\documentclass{article} 
\usepackage{iclr2026_conference,times}

\usepackage{amsmath,amsfonts,bm}

\def\eqref#1{equation~\ref{#1}}

\def\1{\bm{1}}

\DeclareMathAlphabet{\mathsfit}{\encodingdefault}{\sfdefault}{m}{sl}
\SetMathAlphabet{\mathsfit}{bold}{\encodingdefault}{\sfdefault}{bx}{n}

\usepackage{hyperref}
\usepackage{url}
\usepackage{algorithm}
\usepackage{algpseudocode}
\usepackage{amsmath}
\usepackage{booktabs}       
\usepackage{amsfonts}       
\usepackage{nicefrac}       
\usepackage{microtype}      
\usepackage{xcolor}         
\usepackage{graphicx}
\usepackage{wrapfig}
\usepackage{subcaption}
\usepackage{svg}

\title{Follow the MEP: Scalable Neural Representations for Minimum-Energy Path Discovery in Molecular Systems}

\author{%
  Magnus Petersen \\
  Goethe University Frankfurt \& Frankfurt Institute for Advanced Studies\\
  \texttt{mapetersen@fias.uni-frankfurt.de} \\
  \And
    Gemma Roig\\
  Goethe University Frankfurt \& CBMM MIT \& The Hessian Center for AI\\
  \texttt{roig@cs.uni-frankfurt.de}
  \And
  Roberto Covino \\
  Goethe University Frankfurt \& Frankfurt Institute for Advanced Studies\\
  \texttt{covino@fias.uni-frankfurt.de} \\
}

\iclrfinalcopy 
\begin{document}

\maketitle

\begin{abstract}
Characterizing conformational transitions in physical systems remains a fundamental challenge, as traditional sampling methods struggle with the high-dimensional nature of molecular systems and high-energy barriers between stable states. These rare events often represent the most biologically significant processes, yet may require months of continuous simulation to observe. One way to understand the function and mechanics of such systems is through the minimum energy path (MEP), which represents the most probable transition pathway between stable states in the high-friction, low-temperature limit. We present a method that reformulates MEP discovery as a fast and scalable neural optimization problem. By representing paths as implicit neural representations and training with differentiable molecular force fields, our method discovers transition pathways without expensive sampling. Our approach scales to large biomolecular systems through a simple loss function derived from the path's likelihood via the Onsager-Machlup action and a scalable new architecture, AdaPath. We demonstrate this approach on two proteins, including an explicitly hydrated BPTI system with more than 3,500 atoms. Our method identifies a MEP that captures the same conformational change observed in a millisecond-scale molecular dynamics (MD) simulation in just minutes on a standard GPU, rather than weeks on a specialized cluster.
\end{abstract}

\section{Introduction}

When studying the function of large biological systems, molecular dynamics simulations effectively serve as a computational microscope, providing atomic-level detail and control that are impossible with experimental methods alone \citep{dror_biomolecular_2012}. Their true value often lies in not exhaustive sampling of the system's Boltzmann distribution—which is intractable for such systems—but in generating hypotheses from single observed transitions that can then be corroborated through targeted in vitro or in vivo experiments. However, capturing even these individual transition events remains computationally challenging due to the inherent timescale gap between molecular motion and conformational changes. This difficulty arises from transition rates between stable states being exponentially suppressed by the height of the energy barriers that separate them, as described by transition state theory \citep{eyring_activated_1935}. As a result, direct simulation methods spend the vast majority of computational effort sampling already known stable states, making the observation of transitions between them prohibitively expensive, often requiring months of continuous simulation to observe a single biologically significant event.

The search for efficient methods to identify transition pathways between stable states has a rich history in computational physics. Early theoretical work established key principles through the Freidlin-Wentzell \citep{freidlin_random_1998} and Onsager-Machlup \citep{onsager_fluctuations_1953} functionals, which connect the most probable trajectories, including transitions, to minimizers of specific action functionals. Building on these foundations, the minimum-energy path (MEP) between stable molecular states represents the most probable transition pathway in the limit of high friction and low temperature. This path reveals the mechanism of conformational change, identifies key transition states and intermediate configurations, and can thus be highly informative about a system's dynamics. Moreover, the MEP can serve as a scaffold for subsequent sampling of the whole transition path ensemble, making it a powerful tool for understanding molecular dynamics. Methods such as the string method family or chain-of-states approaches \citep{e_string_2002, e_finite_2005, ren_transition_2005, maragliano_string_2006, e_simplified_2007, petersen_teld_2024, dellago_transition_1998} approximate these paths as a discrete series of molecular configurations connected either by artificial spring forces or held together by frequent reparametrization during optimization to ensure continuity. Despite their strong theoretical foundation, these approaches face fundamental challenges stemming from their optimization in Cartesian molecular coordinates \citep{ovchinnikov_free_2011}. The discrete chain of conformations requires either slow periodic reparameterization or a spring force that introduces a hyperparameter to maintain even spacing \citep{lindgren_scaled_2019}. Additional limitations include the difficulty of generating good initial transition guesses \citep{ovchinnikov_free_2011} and limited parallelization, which restricts the scale of practical applications. While this formulation has proven useful for small-scale systems, it has not yet benefited from modern optimization approaches that could address these challenges.
\begin{figure}[H]
\centering
\includegraphics[width=\textwidth]{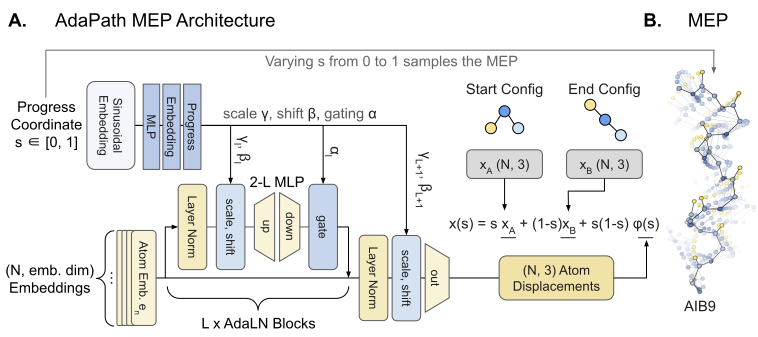}
\caption{Overview of the AdaPath neural architecture and a generated MEP of the 129-atom AIB9 toy system. AdaPath transforms a progress coordinate $s \in [0, 1]$ into molecular configurations through learned atom embeddings processed by shared MLP blocks with adaptive conditioning. The conditioning network generates scale, shift, and gate parameters from $s$ to modulate each block. The final path $\phi(s) \in \mathbb{R}^{3N}$ continuously connects start and end states $x_A$ and $x_B$ via a simple interpolation corrected by the neural network.}
\label{fig:main_overview}
\end{figure}
Recent advances in machine learning have opened the door to strategies addressing this problem. Physics-informed neural networks (PINNs) have shown that neural networks can be trained directly on physical laws expressed as differential equations \citep{raissi_physics-informed_2019}. In addition, the emergence of differentiable molecular dynamics force fields \citep{wang_dmff_2023} has extended this principle to molecular simulations. Together, these developments provide a foundation for deep learning or other optimization approaches that require efficient computation of derivatives beyond forces.

Numerous approaches have emerged over the past decades to tackle the problem of sampling transitions. Traditional simulation-based sampling methods \citep{jung_machine-guided_2023, lazzeri_molecular_2023, bolhuis_transition_2002, dellago_transition_1998, laio_escaping_2002, torrie_nonphysical_1977} rely on expensive molecular dynamics simulations, often requiring initial MD-harvested transition trajectories or prior knowledge of the transition dynamics. More recent machine learning approaches have explored several directions: variational approaches using Doob's h-transform \citep{du_doobs_2024, lee_scaling_2025, das_variational_2019, singh_variational_2023}, stochastic control formulations \citep{yan_learning_2022, holdijk_stochastic_2023}, and even attempts to sample transition trajectories directly using diffusion models \citep{petersen_dynamicsdiffusion_2023, han_geometric_2024, jing_generative_2024}.

Notably, \citet{du_doobs_2024} also uses a neural representation of transition paths but with a different objective, targeting the complete transition path ensemble by additionally parameterizing path width and optimizing based on a framework rooted in Doob's h-transform \citep{doob_conditional_1957} and Wasserstein Lagrangian flows \citep{neklyudov_computational_2024}. Most closely related to our method, \citet{ramakrishnan_implicit_2025} employs a continuous neural parameterization of MEPs inspired by Nudged Elastic Band (NEB) methods, incorporating specialized components for tangential forces, path velocity, and transition state energetics. Another concurrent approach by \citet{raja_action-minimization_2025} leverages pretrained diffusion models as both samplers and force fields. However, these neural approaches face two key limitations: they employ standard MLP architectures that scale poorly with system size or large pretrained models, and they rely on complex multi-term loss functions involving higher-order differential terms known to hamper optimization \citep{rathore_challenges_2024}.

Drawing inspiration from neural implicit representations in computer vision \citep{mildenhall_nerf_2020, sitzmann_implicit_2020}, we aim to address both limitations. We derive a simple, single-term energy-based loss directly from the Onsager-Machlup functional and introduce AdaPath, a novel architecture that employs parameter sharing across atoms. Unlike existing approaches that enforce continuity through complex loss terms, our neural representation achieves smooth paths inherently through architectural design and optimizer choice. This combination enables efficient scaling to large biomolecular systems while maintaining fast, stable optimization. We further perform extensive multi-system benchmarks of our method against classical and deep learning approaches, and find that our approach achieves far lower path and transition state energies. 
\section{Theory and Methods}
\subsection{Derivation of the Energy-Based Loss Function and Neural Implementation}
\label{sec:derivation-of-loss}
To derive our loss function, which, when minimized, yields a continuous curve representing the MEP connecting fixed endpoints $x_A$ and $x_B$, we build upon alternative formulations of the Onsager-Machlup action functional as explored by \citet{vanden-eijnden_geometric_2008} and \citet{olender_yet_1997} in the context of string methods. However, we further simplify the approach after reaching their geometric formulation. A complete derivation with all mathematical details is provided in Appendix~\ref{sec:extended-derivation}. To introduce the Onsager-Machlup action intuitively, we begin with a molecular system evolving under overdamped Langevin dynamics, a model for molecular motion in the high-friction regime characteristic of biomolecular systems:
\begin{align}
  \dot{x}(t) = -\nabla U(x(t)) +\sqrt{2}\eta(t)
\end{align}
$U(x)$ represents the potential energy function, $\eta(t)$ is Gaussian white noise modeling thermal fluctuations, and physical constants are absorbed into the nondimensionalization.
To derive the path probability, we discretize the time interval $[0, T]$ into $N$ steps of duration $\Delta t$, approximating the dynamics as:
\begin{align}
x_{i+1} = x_i - \Delta t\nabla U(x_i) + \sqrt{2\Delta t} \eta_i
\end{align}
Where $\eta_i$ are independent standard Gaussian random variables. The conditional probability of transitioning from $x_i$ to $x_{i+1}$ follows a Gaussian distribution:
\begin{align}
P(x_{i+1}|x_i) \propto \exp\left(-\frac{(x_{i+1} - x_i + \Delta t\nabla U(x_i))^2}{4\Delta t}\right)
\end{align}
Taking the product of these conditional probabilities over all time steps $N$ and then taking the limit as $\Delta t\to 0$ yields the path probability:
\begin{align}
  \mathbb{P}[x(t)] \propto \exp\!(-\frac{1}{4}\int_{0}^{T}\|\dot{x}(t)+\nabla U(x(t))\|^{2}dt)
\end{align}
The exponent corresponds to the Onsager-Machlup action functional \citet{onsager_fluctuations_1953}:
\begin{align}
  S_{\mathrm{OM}}[x]
  = \tfrac{1}{4}\int_{0}^{T}\|\dot{x}(t)+\nabla U(x(t))\|^{2}dt
  \label{eq:som-def}
\end{align}
Thus, maximizing the probability of a transition path (subject to fixed endpoints $x(0)=x_A$ and $x(T)=x_B$) is equivalent to minimizing this action functional. Notably, there are different derivations of the action that yield an additional higher-order derivative term related to trajectory entropy, whose necessity is conditional on the task \citep{adib_stochastic_2008}. Since we are primarily interested in MEPs, we neglect this term. However, the resultant MEPs still empirically align well with the free energy landscape. Notably, this might not be the case for strongly entropy-driven systems such as intrinsically disordered proteins, among others \citep{granata_inverted_2015, caro_entropy_2017, skriver_conformational_2023}. To further simplify, we expand the integrand of the action functional:
\begin{align}
 \|\dot{x}(t) + \nabla U(x(t))\|^{2}
  =
  \|\dot{x}(t)\|^{2}
  +
  2\dot{x}(t)\cdot\nabla U(x(t))
  +
  \|\nabla U(x(t))\|^{2}
\end{align}
For paths with fixed endpoints, the cross-term integrates to a constant difference in potential energy:
\begin{align}
  \int_{0}^{T} \dot{x}(t) \cdot \nabla U(x(t)) dt = U(x_B) - U(x_A)
\end{align}
Since $x_A$ and $x_B$ are fixed in our formulation, this term becomes a constant that does not affect the optimization. We can therefore focus on minimizing the remaining terms. Applying the Cauchy-Schwarz inequality to the remaining terms:
\begin{align}
  \|\dot{x}(t)\|^{2} + \|\nabla U(x(t))\|^{2}
  \ge
  2\|\dot{x}(t)\|\|\nabla U(x(t))\|
\end{align}
This upper bound becomes an equality when $\dot{x}(t)$ is parallel or antiparallel to $\nabla U(x(t))$, with the direction determined by whether the path is ascending or descending the energy landscape. Crucially, we allow the path to optimize its shape and length in our formulation. This freedom enables the path to adjust to follow the energy gradient (or its opposite), making the above upper bound tight at the optimum. When we abandon fixed-time parameterization and allow the path to stretch as needed to align with energy gradients, we can transform the time integral into a purely geometric one by defining the arc-length element $ds_{arc} = \|\dot{x}(t)\|\,dt$ along the spatial curve $\varphi$:
\begin{align}
  \int_{0}^{T}\!\|\nabla U(x(t))\|\|\dot{x}(t)\|dt
  =
  \int_{\varphi}\!\|\nabla U(x)\|ds_{arc}
\end{align}
While the referenced works \citep{vanden-eijnden_geometric_2008, olender_yet_1997} derive and use this geometric action, we further simplify it. This geometric integral represents a natural endpoint for theoretical development for traditional chain-of-states and string methods. In these classical approaches, forces $-\nabla U(x)$ directly update molecular conformations $x$ to minimize $U(x)$, making minimization along a discretized string conceptually straightforward.

However, we want a loss function that acts directly on the neural network parameters $\theta$ in our neural network formulation. While we could parameterize the path using a neural network and then compute forces along the path before backpropagating to our parameters, this would introduce additional gradient computations, which have been noted to lead to optimization challenges in PINNs \citep{rathore_challenges_2024}. We avoid this issue by simplifying to an energy-based loss. 

We represent the transition path as a continuous neural mapping $\varphi_{\theta}(s)$ from a progress coordinate $s\in[0,1]$ to the system's Cartesian coordinates, such that $\varphi_{\theta}(0)=x_A$ and $\varphi_{\theta}(1)=x_B$. We then discretize the path into segments:
\begin{align}
  \int_{\varphi}\!\|\nabla U(x)\|ds_{arc} \approx \sum_{k=0}^{K-2} \|\nabla U(\varphi_k)\|\|\Delta \varphi_k\|
\end{align}
where $\varphi_k = \varphi(s_k)$ and $\Delta \varphi_k = \varphi_{k+1} - \varphi_k$. For a path segment, applying Taylor's theorem, we can express the energy difference as:
\begin{align}
U(\varphi_{k+1}) - U(\varphi_k) &= \nabla U(\varphi_k) \cdot \Delta \varphi_k + O(\|\Delta \varphi_k\|^2) \\
&= \|\nabla U(\varphi_k)\| \|\Delta \varphi_k\| \cos\theta_k + O(\|\Delta \varphi_k\|^2)
\end{align}
Where $\theta_k$ is the angle between the gradient and displacement vectors. For MEPs, the displacement aligns with or against the gradient direction ($\cos\theta_k \to \pm 1$). Critically, this approximation requires $\|\Delta \varphi_k\|$ to be sufficiently small for the higher-order term $O(\|\Delta \varphi_k\|^2)$ to be negligible. This yields:
\begin{align}
\|\nabla U(\varphi_k)\|\|\Delta \varphi_k\| \approx |U(\varphi_{k+1}) - U(\varphi_k)|
\end{align}
This gives us a surrogate objective with the same minimum, provided the Taylor series approximation holds:
\begin{align}
 \sum_{k=0}^{K-2} \|\nabla U(\varphi_k)\|\|\Delta \varphi_k\|
  \approx\sum_{k=0}^{K-2} |U(\varphi_{k+1}) - U(\varphi_k)|
\end{align}
Using the triangle inequality on the absolute energy differences, we can establish that minimizing the simple sum of energies along the path provides an upper bound on this action:
\begin{align}
\sum_{k=0}^{K-2} |U(\varphi_{k+1}) - U(\varphi_k)| \leq \sum_{k=0}^{K-1} U(\varphi_k) - (K-1)\min_k U(\varphi_k) - \frac{U(\varphi_0) + U(\varphi_{K-1})}{2}
\end{align}
Where $\min_k U(\varphi_k)$ is the lowest energy along the path. Since the latter terms are constants during optimization, this leads to our final loss function. To implement this in practice, we sample B points along a parametrized path $\varphi_{\theta}(s)$ and directly minimize the sum of energies at these points, yielding:
\begin{align}
  \mathcal{L}(\theta)
  =
  \frac{1}{B}\sum_{j=1}^{B}
  U(\varphi_{\theta}(s_j))
\end{align}
where $s_j$ are uniformly sampled points in $[0,1]$. While this energy-based loss successfully identifies MEPs when the underlying assumptions hold, it admits a pathological minimum when they break down. Specifically, if $\|\Delta \varphi_k\|$ is large, the validity of the Taylor series approximation breaks down. In that case, the loss can be minimized by paths that remain in one stable state and then rapidly "teleport" to the other, avoiding the high-energy transition region. To prevent this pathological solution, we employ specific architectural choices and optimizer selection to control $\|\Delta \varphi_k\|$ that we detail in the following section.

\subsubsection{AdaPath: A Scalable Neural Architecture for MEPs}

Neural approaches to modeling transitions map $s \in [0,1]$ directly to $\mathbb{R}^{3N}$, requiring hidden layers that scale with system dimensionality, leading to quadratic parameter scaling $O(N^2)$ that limits applicability to large systems.

AdaPath \ref{fig:main_overview} addresses this through parameter sharing across atoms. Each atom $n$ has a learnable embedding $\mathbf{e}_n \in \mathbb{R}^d$, processed through shared transformer-like MLP blocks with adaptive layer norm (adaLN) conditioning \citep{peebles_scalable_2023}. A conditioning network processes sinusoidal embeddings of $s$ to generate scale $\gamma_l$, shift $\beta_l$, and gate $\alpha_l$ parameters that modulate each MLP block $l$, and in addition to a final scale and shift before the final projection layer that projects to the atoms' Cartesian coordinates $\mathbb{R}^{3}$.

Neural implicit representations (NIR) effectively function as smooth lookup tables with a fixed range of inputs rather than complex mappings as in conventional deep learning tasks, allowing neural networks to remain small while embeddings drive the parameter count. In AdaPath, this enables conditioning parameters, which are computed once per $s$ value, to be heavily parameterized, while keeping the main branch operating on each atom embedding lightweight.

AdaPath maintains endpoint constraints through:
\begin{align}
\varphi_\theta(s) = h_0(s)x_A + h_1(s)x_B + b(s) \varphi(s, \{\mathbf{e}_i\}_{i=1}^N)
\end{align}
where $h_0(s) = (1-s)$ and $h_1(s) = s$ provide linear interpolation ensuring $\varphi_\theta(0) = x_A$ and $\varphi_\theta(1) = x_B$, and $b(s) = s(1-s)$ controls network contribution, vanishing at endpoints.

The residual MLP design maintains path continuity and enables gradient flow. While our energy-based loss does not explicitly enforce continuity, residual connections preserve smooth transitions by allowing direct information flow from embeddings to output. However, standard optimizers can cause excessive path stretching around transition regions, see \ref{fig:path_velocity_comparison}. We therefore use the Muon optimizer for small systems \citep{keller_jordan_muon_2024}, as it maintains lower Lipschitz constants \citep{large_scalable_2024}, ensuring adequate path presence in transition regions for accurate energy estimates. For large systems, Adam \citep{kingma_adam_2017} suffices; we posit that this might be due to neural networks' increasing inability to model discontinuities as dimensionality increases \citep{petersen_optimal_2018}. We further scale the energy function and use gradient clipping, as outlined in \ref{sec:energyspikes}, to avoid high energy spikes caused by steric clashes between atoms.
\subsection{Established Approaches to MEP Finding}

All MEP discovery methods must solve a fundamental optimization challenge: finding paths that simultaneously minimize energy while maintaining smooth continuity between endpoint configurations. This dual objective creates an inherent tension; energy minimization alone may produce discontinuous jumps between low-energy regions, while continuity enforcement alone may force paths through high-energy barriers. Different methodological families resolve this through distinct mathematical formulations, each with characteristic advantages and limitations that our neural approach addresses. The approaches discussed below represent broad families with numerous variations, extensively cataloged in reviews such as \citet{e_transition-path_2010-1}.

\subsubsection{Explicit Continuity Penalties}

The chain-of-states family \citep{pratt_statistical_1986, elber_method_1987, ulitsky_new_1990} handles the energy-continuity trade-off through composite objective functions that explicitly balance both terms. Given a sequence $\{x_0,\ldots,x_N\}$ with $x_0=x_A$ and $x_N=x_B$, the optimization targets:
\begin{align}
\dot{x}_i = -\nabla U(x_i) + k \frac{x_{i+1}+x_{i-1}-2x_{i}}{\Delta s}
\end{align}
Where the first term enforces energy minimization and the spring term constant $k$ penalizes discontinuities. However, this formulation suffers from "corner cutting", where the path takes shortcuts through high-energy regions when spring forces dominate.

The Nudged Elastic Band (NEB) method \citep{jonsson_nudged_1998, henkelman_climbing_2000-1} refines this approach by projecting spring forces onto the path normal direction, yielding:
\begin{align}
\dot{x}_i = -[\nabla U(x_i)]^\perp + k \frac{x_{i+1}+x_{i-1}-2x_{i}}{\Delta s}_\parallel
\end{align}
Where $[\nabla U]^\perp$ represents forces perpendicular to the path and the spring term acts only parallel $\parallel$ to it. This decoupling reduces corner cutting but requires careful force projection and still faces scaling challenges in high-dimensional systems, often leading to jagged paths \citep{zhang_free-end_2016}.

\subsubsection{Continuity via Geometric Constraints}

The string method \citep{e_string_2002, e_finite_2005, ren_transition_2005, maragliano_string_2006} takes an alternative approach, separating energy minimization from continuity enforcement through alternating optimization. The method seeks a curve $\varphi(s)$ satisfying the orthogonality condition:
\begin{align}
[\nabla U]^\perp = \nabla U - (\nabla U \cdot \hat{\tau})\hat{\tau} = 0
\end{align}
where $\hat{\tau} = \varphi'/|\varphi'|$ is the unit tangent vector along the string. This orthogonality condition has an intuitive interpretation: at each point along the MEP, the forces acting on the system are entirely parallel to the path itself, with no perpendicular components trying to push the system away from the path. The simplest dynamic for evolving toward this MEP is the following:
\begin{align}
\frac{\partial \varphi}{\partial t} = -[\nabla U(\varphi)]^\perp + \lambda \hat{\tau}
\end{align}
where the Lagrange multiplier $\lambda \hat{\tau}$ enforces parameterization constraints. However, in practice, implementations alternate between: (1) evolving images $x_i$ along the string using energy gradients $\dot{x}_i = -\nabla U(x_i)$, and (2) reparameterization to maintain equal spacing:
\begin{align}
s_i = \frac{\sum_{j=0}^i |x_{j+1} -x_j|}{\sum_{j=0}^{N-1} |x_{j+1} - x_j|}
\end{align}
This effectively enforces the same constraint as the Lagrange multiplier \citep{e_transition-path_2010-1}.
\subsubsection{Computational Limitations}

These classical approaches share fundamental optimization limitations stemming from their discrete optimization in Cartesian molecular coordinates \citep{ovchinnikov_free_2011}. Chain-of-states methods require careful tuning of the spring constant and face corner cutting, while string methods demand complex reparameterization procedures \citep{lindgren_scaled_2019}. Both struggle with initial transition guesses \citep{ovchinnikov_free_2011}, local minima trapping, and limited parallelization that restricts applications to smaller systems, as each image is separately parameterized. Our neural representation addresses these challenges by encoding continuity architecturally, simplifying optimization to a single energy-based objective while leveraging optimization in neural parameter space to reduce local minima susceptibility.
\section{Experiments}
\begin{wrapfigure}{R}{0.49\textwidth}
  \vspace{-15pt}
  \centering
  \includegraphics[width=\linewidth]{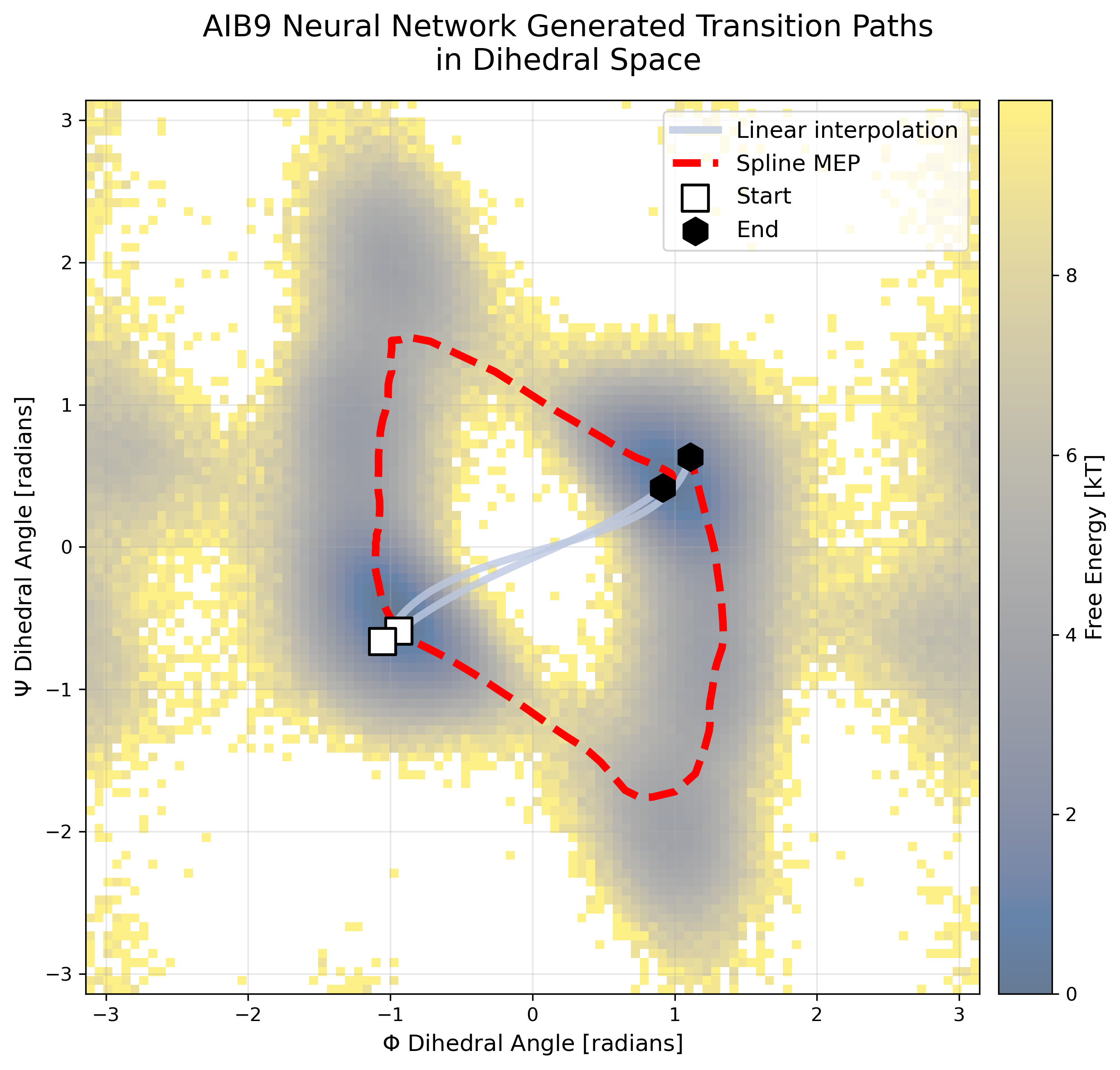}
  \caption{AIB9 free energy surface projected onto central residue dihedral angles ($\phi$, $\psi$). Two distinct MEPs were found with different random seeds, demonstrating the method's ability to find multiple distinct transition pathways.}
  \label{fig:aib9_landscape}
  \vspace{-40pt}
\end{wrapfigure}
We evaluate our method on two molecular systems: the AIB9 peptide and the Bovine Pancreatic Trypsin Inhibitor (BPTI). These systems represent different scales of complexity and serve to validate our method's ability to discover physically meaningful transition paths. In Appendix \ref{sec:bench_ala}, \ref{sec:bench_aib9} and \ref{sec:bench_bpti}, we show extensive benchmarks of our method with the related approaches outlined in \citep{du_doobs_2024, lee_scaling_2025} and classical MEP approaches, namely chain-of-states, NEB, and the string method, and demonstrate that we outperform all other methods in terms of both mean and maximum path energy. In Appendix \ref{sec:ablation}, we present ablation studies analyzing the impact of our architectural choices on MEP quality.
\subsection{AIB9 Transition Path Discovery}
We first examine the AIB9 system, a 9-residue artificial protein with 129 atoms that exhibits two well-defined metastable states, making it an ideal test case for transition path methods. It is small enough for extensive reference simulations yet complex enough to exhibit realistic conformational changes, and its transitions are easily visualized through dihedral angle projections. We therefore also use it for the majority of benchmarking and ablation experiments. Unlike artificial toy problems, AIB9 allows direct validation with actual biomolecular force fields. We use the AMBER ff15ipq-m force field for protein-mimetics \citep{bogetti_twist_2020} implemented in DMFF \citep{wang_dmff_2023}, and we generate multiple MEPs between the system's two metastable states.

Figure \ref{fig:aib9_landscape} shows the discovered transition paths projected onto the central residue's $\phi$-$\psi$ dihedral angle space. The method identifies multiple physically plausible pathways between the states, which are also observed with long MD simulations. For each path, different conformations in the states were picked as start and end points; this, along with the neural network's random initialization, leads to the discovery of multiple different transitions. Notably, the discovered MEPs align well with the free energy landscape despite our choice of an action functional without the trajectory entropy term and the assumption of overdamped dynamics, which may not be ideal for this vacuum system, but is less problematic for hydrated systems.

\section{BPTI Conformational Change Pathway}
\label{sec:BPTI}
Bovine Pancreatic Trypsin Inhibitor (BPTI) is a clinically significant protein used as an anticoagulant in medical procedures. BPTI is 58 residues long (892 atoms) and exhibits complex conformational changes involving disulfide bond rearrangements paired with backbone changes.

To investigate BPTI's conformational transitions, we use as reference structures five states from a landmark molecular dynamics (MD) simulation performed by D.E. Shaw Research on their Anton supercomputer \citep{shaw_atomic-level_2010}. These publicly available snapshots provide rare benchmark structures for validating our approach to BPTI's transition mechanisms, making it an excellent system to test the method on a larger scale. Full details of our BPTI system preparation and hydration are provided in Appendix~\ref{sec:bpti_setup}.

\begin{figure}[h]
  \vspace{-5pt}
\centering
\begin{subfigure}{0.49\textwidth}
    \centering
    \includegraphics[height=2.79in]{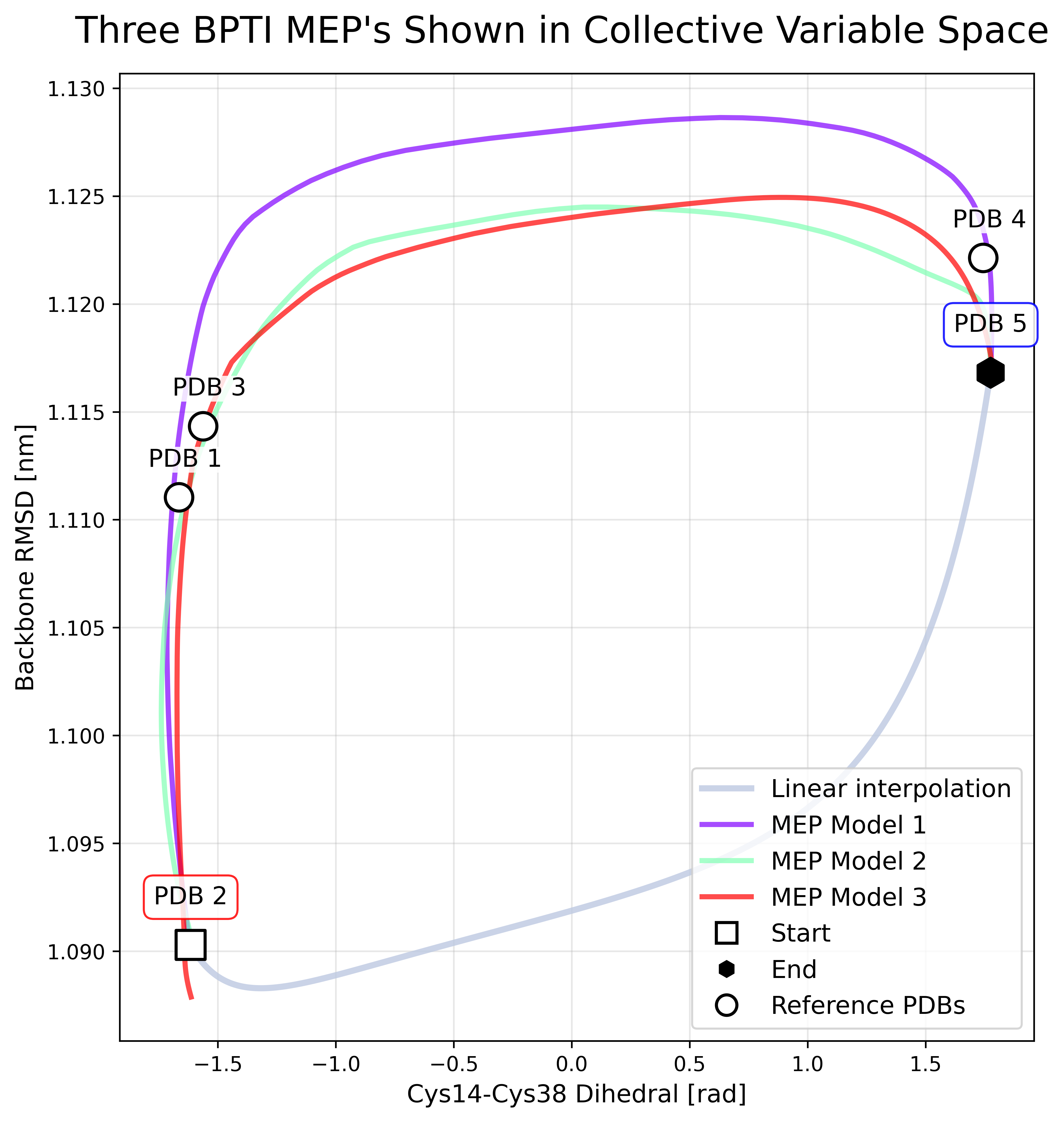}
\end{subfigure}
\hfill
\begin{subfigure}{0.49\textwidth}
    \centering
    \includegraphics[height=2.8in]{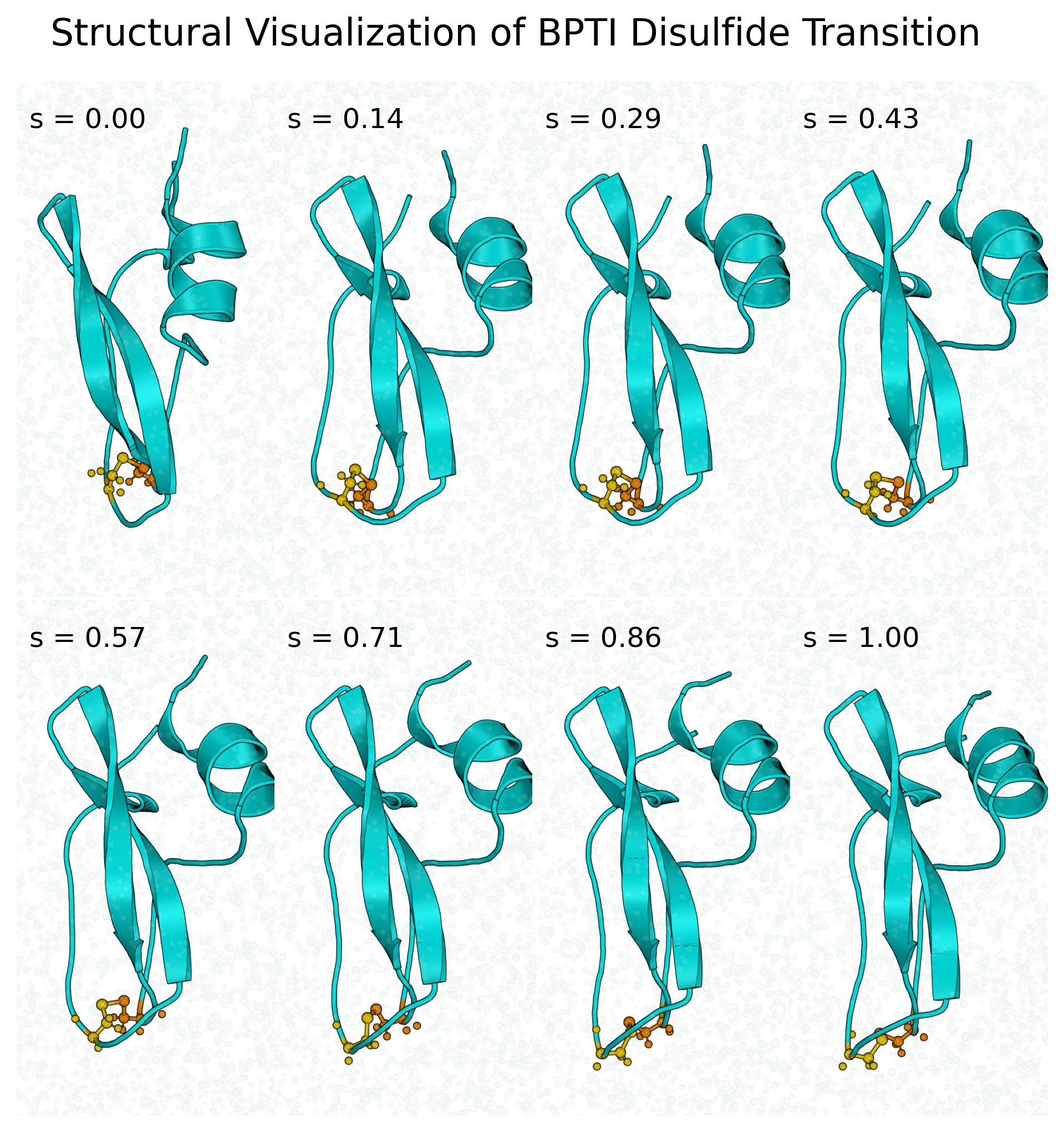}
\end{subfigure}
\caption{Analysis of our discovered BPTI transition pathway. \textbf{Left:} Evolution of BPTI dynamics projected onto two collective variables matching the original D.E. Shaw study: the Cys14-Cys38 disulfide torsion angle (formed by CB14–SG14–SG38–CB38 atoms) versus the mean-centered backbone RMSD of residues 4-54. Reference structures from the D.E. Shaw simulation are marked with circles. Our optimized paths pass through all intermediate snapshots, demonstrating that it discovers the same transition mechanism observed in the millisecond-scale MD simulation. \textbf{Right:} Structural visualization of the BPTI and water system highlighting the Cys14-Cys38 disulfide bond conformation along one of our optimized transition paths. Each panel represents a different value of the progress coordinate $s$, showing how the disulfide bond rotates during the transition. This rotation is a key feature of the BPTI conformational change. For clarity, the protein is shown using its secondary structure representation except for the disulfide bond, and the water molecules are rendered semi-translucent.}
\label{fig:bpti_analysis}
\vspace{-10pt}
\end{figure}

We generate multiple transition MEPs connecting two BPTI states from the D.E. Shaw simulation. The intermediate snapshots from the D.E. Shaw simulation serve as validation points; they were not used as inputs but represent physically realized conformations from the whole trajectory. To analyze this path, we compute two collective variables matching those used in the original D.E. Shaw study: (1) the disulfide torsion angle between Cys14 and Cys38, calculated using the dihedral angle formed by CB14–SG14–SG38–CB38 atoms, and (2) the backbone RMSD, root mean square displacement, of residues 4-54 after mean-centering. Remarkably, as shown in Figure \ref{fig:bpti_analysis}, our optimized path passes through all intermediate snapshots, indicating that it successfully discovers the same transition mechanism as the computationally intensive MD simulation.

We matched their simulation conditions by incorporating explicit water molecules to ensure a meaningful comparison with the Anton simulation. We hydrate both the initial and final BPTI conformations, as done in a conventional MD system setup. This increases the number of atoms from 892 to 3,500. To handle water molecules during transitions, we pair proximal water molecules between start and end states using the Hungarian algorithm \citep{kuhn_hungarian_1955}.

The system was modeled using the AMBER99sb-ILDN force field \citep{lindorff-larsen_improved_2010} with the TIP3P water model \citep{jorgensen_comparison_1983, neria_simulation_1996}. Each MEP was computed in approximately 15 minutes on a single A6000 GPU, requiring only $\sim192,000$ force field evaluations compared to the D.E. Shaw simulation, which conducted approximately 412 billion force field evaluations to generate the 1.03 millisecond trajectory capturing the same conformational changes. This represents a six orders of magnitude reduction in computational effort. 

While our approach does not provide the same statistical ensemble information as the reference MD, it efficiently captures the key transition pathway. It is worth noting that even the reference millisecond-scale simulation has not achieved equilibrium sampling, as it is comprised of a single transition. Appendix \ref{sec:performance} provides a detailed computational performance comparison.

\section{Comparative Evaluation and Ablation Studies}
To systematically validate our approach, we conducted extensive benchmarks against established methods and ablation studies examining our architectural choices across three molecular systems of increasing complexity. We evaluate path quality using two key metrics: the maximum energy along the path (indicating transition state barriers), the arc-length adjusted mean energy (accounting for how quickly paths traverse high-energy regions), and the standard deviation of this quantity for non-deterministic methods. The full results for the benchmarks are presented in sections \ref{sec:bench_ala}, \ref{sec:bench_aib9}, and \ref{sec:bench_bpti}, and the ablations are discussed in section \ref{sec:ablation} of the appendix, respectively.

\textbf{Benchmark Results:} AdaPath consistently outperforms classical approaches (Chain-of-States, String Method, NEB) and deep learning approaches like Doob's Lagrangian in both computational efficiency and path quality. For the alanine dipeptide, we achieve superior energy barriers (-86.6 vs. -71.1 kJ/mol for Chain-of-States) while requiring $3.8\times$ fewer force evaluations (6.4e4 vs. 2.4e5). Here and on other systems, Chain-of-States emerged as the best-performing classical method. We note that Doob's Lagrangian targets the more ambitious objective of modeling the entire transition path ensemble rather than discovering a single MEP, making direct performance comparisons inherently limited.

This performance advantage scales to larger systems: on AIB9 (129 atoms), AdaPath reaches -644 kJ/mol maximum energy compared to Chain-of-States' -461 kJ/mol and substantially outperforms all other tested methods (Appendix~\ref{sec:bench_aib9}). On the BPTI system with explicit solvent (3,500 atoms), classical methods either fail to converge and produce erratic paths that bypass intermediate reference structures from the landmark D.E. Shaw simulation, as shown qualitatively in Appendix~\ref{sec:bench_bpti}.

\textbf{Architectural Validation:} Ablation studies confirm our core design choices (Appendix~\ref{sec:ablation}). Parameter sharing proves essential for scalability; standard MLPs fail catastrophically on large systems, with even billion-parameter variants showing poor performance on BPTI. Our optimizer analysis reveals that Muon maintains stable path continuity, while Adam exhibits erratic "teleportation" behavior that rapidly jumps between states without traversing transition regions in small systems. The logarithmic energy transformation proves critical for training stability, reducing energy variance by orders of magnitude.

\section{Limitations}
While our approach demonstrates strong performance on protein systems up to 3,500 atoms, several limitations should be noted. First, our method is constrained to systems with suitable differentiable force fields, though this limitation is diminishing as more implementations become available. Second, our method produces single representative MEPs rather than sampling the complete statistical ensemble of transition pathways. While this represents a trade-off between computational efficiency and exhaustive sampling, the mechanistic insights provided are often the primary objective for practitioners studying biomolecular transitions.
\section{Conclusion}
We demonstrated that reformulating MEP discovery as a continuous neural optimization problem, combined with a scalable neural architecture, enables efficient discovery of molecular transition mechanisms for large biomolecular systems in explicitly modeled solvent. Our approach makes two key contributions: deriving a simple energy-based loss function directly from the Onsager-Machlup action functional, and introducing AdaPath. This architecture scales favorably with system size via parameter sharing across the atom dimension. By encoding path continuity through architectural design rather than explicit loss terms, we sidestep known optimization challenges in PINNs and multi-loss training.
Several promising directions for future development emerge from this work. The scaling properties of AdaPath suggest the method could be extended to even larger biomolecular systems, such as membrane proteins or multi-domain complexes that routinely contain tens of thousands of atoms in MD simulations, not counting water. Our discovered MEPs could serve as guidance paths for enhanced sampling methods, enabling free energy calculations through escorted Jarzynski-like estimators \citep{jarzynski_nonequilibrium_1997, vaikuntanathan_escorted_2011} by using the MEP as a basis for a non-equilibrium pulling protocol between stable states. The architecture and loss function may also prove valuable for other non-biological physical systems where rare events might be of interest.
\bibliography{references}
\bibliographystyle{iclr2026_conference}

\appendix
\section{Appendix}
\label{sec:appendix}
\subsection{Extended Derivation of the energy-based Loss Function}
\label{sec:extended-derivation}
In this appendix, we provide a comprehensive derivation of our energy-based loss function for discovering minimum-energy paths (MEPs) between molecular configurations. The derivation proceeds in four main steps: (1) establishing the path probability density from overdamped Langevin dynamics, (2) reformulating this probability in terms of the Onsager-Machlup action functional, (3) simplifying this functional to a geometric line integral, and (4) developing a discretized approximation suitable for neural network optimization.

Our goal is to find the most probable transition path between two stable molecular configurations $x_A$ and $x_B$. This path corresponds to the minimum-energy pathway through the potential energy landscape and represents the most likely mechanism of conformational change.

\subsubsection*{Assumptions and Approximations}

The derivation relies on several key assumptions:
\begin{itemize}
    \item The molecular system evolves according to overdamped Langevin dynamics, appropriate for high-friction biomolecular environments
    \item The path endpoints $x_A$ and $x_B$ are fixed and correspond to stable states
    \item The potential energy function $U(x)$ is differentiable everywhere along the path
    \item We can represent the path as a continuous, parameterized curve in configuration space with sufficient flexibility
    \item The path's parameterization allows for effective length changes, permitting the path to adjust its stretching as needed to follow energy gradients optimally
\end{itemize}

We also make several important approximations during the derivation:
\begin{itemize}
    \item We use a first-order Taylor expansion to relate energy differences to gradient magnitudes, which requires displacement vectors $\|\Delta \varphi_k\|$ to be sufficiently small
    \item We introduce stochastic sampling to estimate the continuous energy integral
    \item For a MEP, the displacement tends to align either parallel or antiparallel with the gradient direction (i.e., $\cos\theta_k \to \pm 1$ for path segments)
    \item We use the triangle inequality to establish an upper bound on the sum of absolute energy differences
\end{itemize}

While the last two approximations do not change the location of the global minimum, they do affect the optimization landscape before that minimum is reached. Critically, when the Taylor approximation breaks down due to large displacement vectors, the resulting energy-based loss can admit pathological solutions where paths remain in stable states and then rapidly "teleport" to avoid high-energy transition regions.

\subsubsection{From the Onsager-Machlup Action to a geometric Line Integral}

We begin with a molecular system evolving under overdamped Langevin dynamics, a standard model for high-friction molecular motion:

\begin{align}
  \dot{x}(t) = -\nabla U(x(t)) + \sqrt{2}\eta(t)
\end{align}
where $x(t) \in \mathbb{R}^{3N}$ represents the molecular configuration, $U(x)$ is the potential energy function, and $\eta(t)$ is a Gaussian white noise with $\langle\eta_i(t)\eta_j(t')\rangle = \delta_{ij}\delta(t-t')$.

To derive the path probability, we discretize the time interval $[0, T]$ into $N$ steps of duration $\Delta t$, approximating the dynamics as:
\begin{align}
x_{i+1} = x_i - \Delta t\nabla U(x_i) + \sqrt{2\Delta t} \eta_i
\end{align}
where $\eta_i$ are independent standard Gaussian random variables with zero mean and unit variance. 

The conditional probability of transitioning from $x_i$ to $x_{i+1}$ follows a Gaussian distribution:
\begin{align}
P(x_{i+1}|x_i) &= \frac{1}{\sqrt{4\pi\Delta t}^{d}} \exp\left(-\frac{(x_{i+1} - x_i + \Delta t\nabla U(x_i))^2}{4\Delta t}\right) \\
&\propto \exp\left(-\frac{(x_{i+1} - x_i + \Delta t\nabla U(x_i))^2}{4\Delta t}\right)
\end{align}
where $d$ is the dimensionality of the system. The probability of the entire discretized path $\{x_0, x_1, \ldots, x_N\}$ is given by the product of these conditional probabilities:
\begin{align}
P(x_0, x_1, \ldots, x_N) &= P(x_0)\prod_{i=0}^{N-1}P(x_{i+1}|x_i) \\
&\propto \prod_{i=0}^{N-1}\exp\left(-\frac{(x_{i+1} - x_i + \Delta t\nabla U(x_i))^2}{4\Delta t}\right) \\
&= \exp\left(-\sum_{i=0}^{N-1}\frac{(x_{i+1} - x_i + \Delta t\nabla U(x_i))^2}{4\Delta t}\right)
\end{align}

Taking the limit as $\Delta t \to 0$ and $N \to \infty$ with $N\Delta t = T$ fixed, this sum approaches a path integral:
\begin{align}
\lim_{\substack{\Delta t \to 0 \\ N\Delta t = T}} \sum_{i=0}^{N-1}\frac{(x_{i+1} - x_i + \Delta t\nabla U(x_i))^2}{4\Delta t} 
= \frac{1}{4}\int_{0}^{T}\|\dot{x}(t)+\nabla U(x(t))\|^{2}dt
\end{align}

Thus, the path probability can be written as:
\begin{align}
  \mathbb{P}[x(t)] \propto \exp\!(-\frac{1}{4}\int_{0}^{T}\|\dot{x}(t)+\nabla U(x(t))\|^{2}dt)
\end{align}

The exponent term is recognized as the Onsager-Machlup action functional:
\begin{align}
  S_{\mathrm{OM}}[x]
  = \frac{1}{4} \int_{0}^{T} \|\dot{x}(t) + \nabla U(x(t))\|^2 dt
\end{align}

This action functional penalizes paths that deviate from following the force $-\nabla U(x)$. Maximizing the probability of a transition path (subject to fixed endpoints $x(0)=x_A$ and $x(T)=x_B$) is equivalent to minimizing this action functional.

Expanding the integrand:
\begin{align}
  \|\dot{x}(t) + \nabla U(x(t))\|^2 = \|\dot{x}(t)\|^2 + 2\dot{x}(t) \cdot \nabla U(x(t)) + \|\nabla U(x(t))\|^2
\end{align}

For paths with fixed endpoints $x(0) = x_A$ and $x(T) = x_B$, the cross-term integrates to a constant difference in potential energy:
\begin{align}
  \int_{0}^{T} \dot{x}(t) \cdot \nabla U(x(t)) dt = \int_{0}^{T} \frac{d}{dt}U(x(t)) dt = U(x_B) - U(x_A)
\end{align}

We can define a modified action functional that, when minimized, is equivalent to minimizing the original Onsager-Machlup functional for fixed endpoints:
\begin{align}
  \tilde{S}[x] = \int_{0}^{T} (\|\dot{x}(t)\|^2 + \|\nabla U(x(t))\|^2) dt
\end{align}

By the Cauchy-Schwarz inequality:
\begin{align}
  \|\dot{x}(t)\|^2 + \|\nabla U(x(t))\|^2 \geq 2\|\dot{x}(t)\|\|\nabla U(x(t))\|
\end{align}
With equality when $\dot{x}(t)$ is parallel or antiparallel to $\nabla U(x(t))$, with the direction determined by whether the path is ascending or descending the energy landscape. This is precisely the defining characteristic of the MEP - a path that follows the potential energy landscape's gradient while avoiding high-energy regions.

When this equality holds, our action simplifies to:
\begin{align}
  \tilde{S}[x] = \int_{0}^{T} 2\|\dot{x}(t)\|\|\nabla U(x(t))\| dt
\end{align}

The term $\|\dot{x}(t)\|dt$ represents the infinitesimal arc length element $ds_{\text{arc}}$ along the path, allowing us to rewrite the action as a line integral:
\begin{align}
  S_{\mathrm{geo}}[x] = 2 \int_{\varphi} \|\nabla U(x)\| ds_{\text{arc}}
\end{align}
where $\varphi$ is the path in configuration space. This geometric form is independent of the path's parameterization and depends solely on the potential energy landscape.

\subsubsection{Discretization and Connection to Energy Differences}

To compute this geometric action numerically, we introduce a progress coordinate $s \in [0,1]$ and a mapping $\varphi_\theta(s)$ such that $\varphi_\theta(0)=x_A$ and $\varphi_\theta(1)=x_B$, where $\theta$ represents the network parameters. The relationship between the progress coordinate and arc length is:
\begin{align}
  ds_{\text{arc}} = \left\|\frac{d\varphi_\theta}{ds}\right\| ds
\end{align}

We discretize the path with $K$ points, setting $s_k = \frac{k}{K-1}$ and $\varphi_k = \varphi_\theta(s_k)$ for $k = 0, 1, ..., K-1$. The discretized geometric action becomes:
\begin{align}
  S_{\mathrm{geo}}[\varphi_\theta] \approx 2 \sum_{k=0}^{K-2} \|\nabla U(\varphi_k)\| \cdot \|\Delta \varphi_k\|
\end{align}
where $\Delta \varphi_k = \varphi_{k+1} - \varphi_k$.

Using Taylor's theorem, we can relate energy differences to gradient magnitudes:
\begin{align}
  U(\varphi_{k+1}) - U(\varphi_k) &= \nabla U(\varphi_k) \cdot \Delta \varphi_k + O(\|\Delta \varphi_k\|^2) \\
  &= \|\nabla U(\varphi_k)\| \|\Delta \varphi_k\| \cos \theta_k + O(\|\Delta \varphi_k\|^2)
\end{align}
where $\theta_k$ is the angle between the gradient and displacement vectors. 

Critically, this Taylor expansion approximation is only valid when $\|\Delta \varphi_k\|$ is sufficiently small for the higher-order term $O(\|\Delta \varphi_k\|^2)$ to be negligible. When displacement vectors become large, the approximation breaks down. It can lead to pathological solutions, where the optimization finds paths that avoid high-energy transition regions by rapidly "teleporting" between stable states, rather than discovering smooth, minimum-energy routes.

For a path following the minimum-energy trajectory, the gradient aligns with or against the path direction ($\cos \theta_k \to \pm 1$ as $\Delta s \to 0$, depending on whether the path is ascending or descending the energy landscape), giving at the minimum of the optimization:
\begin{align}
  |U(\varphi_{k+1}) - U(\varphi_k)| \approx \|\nabla U(\varphi_k)\| \cdot \|\Delta \varphi_k\|
\end{align}

This yields an approximation of the geometric action:
\begin{align}
  S_{\mathrm{geo}}[\varphi_\theta] \approx 2 \sum_{k=0}^{K-2} |U(\varphi_{k+1}) - U(\varphi_k)|
\end{align}

This forms a surrogate objective with the same minimum as our original action, provided the Taylor approximation remains valid throughout optimization.

\subsubsection{Establishing the energy-based Loss Function}

We need to bound the sum of absolute energy differences to establish a direct connection to an energy-based loss function. Using the triangle inequality, for any two points and any constant $c$:
\begin{align}
  |U(\varphi_{k+1}) - U(\varphi_k)| \leq \frac{(U(\varphi_{k+1}) - c) + (U(\varphi_k) - c)}{2} = \frac{U(\varphi_{k+1}) + U(\varphi_k) - 2c}{2}
\end{align}

Choosing $c = \min_{s \in [0,1]} U(\varphi_\theta(s))$ to be the minimum energy along the entire path and summing over all segments:
\begin{align}
  \sum_{k=0}^{K-2} |U(\varphi_{k+1}) - U(\varphi_k)| &\leq \sum_{k=0}^{K-2} \frac{U(\varphi_{k+1}) + U(\varphi_k) - 2c}{2} \\
  &= \frac{1}{2}\left(\sum_{k=0}^{K-2} U(\varphi_{k+1}) + \sum_{k=0}^{K-2} U(\varphi_k) - 2(K-1)c\right)
\end{align}

We can rewrite the first sum as $\sum_{k=1}^{K-1} U(\varphi_k)$ and the second as $\sum_{k=0}^{K-2} U(\varphi_k)$. Combining these terms:
\begin{align}
  \sum_{k=0}^{K-2} |U(\varphi_{k+1}) - U(\varphi_k)| &\leq \frac{1}{2}\left(\sum_{k=1}^{K-1} U(\varphi_k) + \sum_{k=0}^{K-2} U(\varphi_k) - 2(K-1)c\right) \\
  &= \frac{1}{2}\left(\sum_{k=0}^{K-1} U(\varphi_k) - U(\varphi_0) + \sum_{k=0}^{K-1} U(\varphi_k) - U(\varphi_{K-1}) - 2(K-1)c\right) \\
  &= \sum_{k=0}^{K-1} U(\varphi_k) - \frac{U(\varphi_0) + U(\varphi_{K-1})}{2} - (K-1)c
\end{align}

Since $c$, $U(\varphi_0)$, and $U(\varphi_{K-1})$ are constants during optimization, minimizing $\sum_{k=0}^{K-1} U(\varphi_k)$ is equivalent to minimizing the upper bound on the geometric action. This provides the theoretical justification for our energy-based loss function.

\subsubsection{Neural Network Implementation}

For practical optimization, we parametrize the path using a neural network $\varphi_{\theta}(s)$ with parameters $\theta$. To implement our approach, we sample points along the path and minimize the energy at these sampled locations. This represents a transition from the theoretical derivation using discrete segments to a practical implementation using sampled points for more efficient optimization. We employ stochastic sampling of the progress coordinate to construct our loss function:
\begin{align}
  \mathcal{L}(\theta) = \frac{1}{B} \sum_{j=1}^{B} U(\varphi_{\theta}(s_j))
\end{align}
where $s_j \sim \mathcal{U}[0,1]$ are uniformly sampled progress coordinates and $B$ is the batch size (the number of points sampled along the path during each optimization step).

This stochastic approach provides an unbiased estimator of the expected energy along the path:
\begin{align}
  \mathbb{E}_{s \sim \mathcal{U}[0,1]}[U(\varphi_{\theta}(s))] = \int_{0}^{1} U(\varphi_{\theta}(s)) ds
\end{align}

By minimizing $\mathcal{L}(\theta)$ through gradient-based optimization, the neural network learns to represent low-energy pathways connecting the stable states, effectively discovering the MEP predicted by the Onsager-Machlup framework. The resulting path follows the potential energy landscape's gradient while avoiding high-energy regions, representing the most probable transition mechanism in the overdamped limit. However, as noted above, the validity of this approach depends critically on maintaining sufficiently small displacement vectors $\|\Delta \varphi_k\|$ during optimization to ensure the underlying Taylor approximation remains valid.

\section{Benchmarks: Alanine Dipeptide}
\label{sec:bench_ala}
We provide a direct comparison between our AdaPath method, \citet{du_doobs_2024}, \cite{lee_scaling_2025}, as well as classical, well-established methods like the string method, chain-of-states, and nudged elastic band, in terms of computational efficiency, as evaluated by the number of force field/energy function calls, and the maximum energies observed along the paths, corresponding to the peak of the potential barrier, both as a mean across five seeds as well as the lowest value achieved across all runs. We further compute the arc-length-adjusted mean energy along the path. We chose this metric as naive averaging along a discretized path does not account for how fast a path is traversing a high-energy region; it could, therefore, shortcut a region of high energy but have a very low non-adjusted energy. To provide a rigorous comparison, we conducted benchmarks using experimental conditions identical to those reported by Du et al. for alanine dipeptide: AMBER14 force field, 300 K temperature, and also referred to their results when applicable, but adjusted the significant figure notation. Table \ref{tab:benchmark} presents these results alongside those reported in \cite{du_doobs_2024} and their recent follow-up work \cite{lee_scaling_2025}. The only exception is that we do not evaluate trajectory probability, as the assumption of optimization over a non-physical path length makes calculating this quantity nonsensical. Furthermore, the arc-length-adjusted energy quantities are not applicable to Doob's method, as they do not model a single path but rather the entire transition path ensemble.

For the classical methods, we conducted extensive hyperparameter sweeps to ensure fair comparison and optimal performance. Chain-of-States was optimized across path discretizations (20-60 images), spring constants (1e-7 to 1e-1), and learning rates (5e-2 to 1e-1), with the best configuration using 40 images, a spring constant of 1e-6, and the Adam optimizer with a learning rate of 5e-2. The String Method was evaluated across image counts (20-60), timesteps (0.1-4000), and reparameterization frequencies (1-10 steps), achieving optimal performance with 40 images, timestep 3900, and reparameterization every three steps. NEB was tested across various path discretizations (20-60 images), spring constants (1e-7 to 1e-1), and learning rates (1e-6 to 1e-1), with the best results obtained using 60 images, a spring constant of 2e-5, and the Adam optimizer with a learning rate of 1e-5. These configurations represent the optimal settings identified through systematic exploration of the hyperparameter space for each method. All classical approaches we study are deterministic, and hence are only run once, unlike the stochastic approaches, which are run five times.

Since the field lacks standardized benchmarks, we believe that building on the work from \citep{du_doobs_2024} could be a helpful step toward having a more standardized approach to methods aiming to solve for MEPs, the transition path ensemble, or other similar approaches. We want to reiterate that \citep{du_doobs_2024} is more ambitious in its sampling objectives, and therefore, one-to-one comparisons are not entirely appropriate. However, the transition state energy, as represented by the Max Energy entry, should be consistent between MEP approaches and methods that aim to sample the entire transition ensemble.

\begin{table}[h]
\centering
\caption{Comprehensive benchmark comparison of transition path methods on alanine dipeptide system. Results include computational efficiency (measured by the total number of force field/energy function evaluations), transition state energies (in kJ/mol), and path quality metrics. Classical methods (Chain-of-States, String Method, NEB) were optimized through extensive hyperparameter sweeps. Doob's Lagrangian results are reproduced from \cite{du_doobs_2024} and \cite{lee_scaling_2025}. Max Energy represents the mean $\pm$ standard deviation across five seeds of the highest energy barrier along discovered paths across multiple runs. MinMax Energy shows the lowest maximum energy achieved in any single run. Mean Energy is the arc-length-adjusted average energy along the path (kJ/mol).}
\label{tab:benchmark}
\begin{tabular}{lrrrr}
\toprule
\textbf{Method} & \textbf{\# Evals} & \textbf{Max Energy} & \textbf{MinMax Energy} & \textbf{Mean Energy} \\
\midrule
MCMC variable length & 2.1e7 & $740 \pm 700$ & $52.4$ & N/A \\
MCMC* & 1.3e9 & $288 \pm 128$ & $60.5$ & N/A \\
\midrule
MCMC variable length & 1.9e8 & $413 \pm 335$ & $27.0$ & N/A \\
MCMC & $>$ 1e10 & N/A & N/A & N/A \\
\midrule
MCMC variable length & $>$ 1e10 & N/A & N/A & N/A \\
MCMC & $>$ 1e10 & N/A & N/A & N/A \\
Doob's Cartesian & 3.8e7 & $726.40 \pm 0.07$ & $726.2$ & N/A \\
w/ 2 Mixtures & 5.1e7 & $709 \pm 162$ & $513.7$ & N/A \\
w/ 5 Mixtures & 5.1e7 & $541 \pm 278$ & $248.0$ & N/A \\
Doob's Internal & 3.8e7 & $-14.62 \pm 0.02$ & $-14.67$ & N/A \\
w/ 2 Mixtures & 5.1e7 & $-15.38 \pm 0.14$ & $-15.54$ & N/A \\
w/ 5 Mixtures & 5.1e7 & $-15.5 \pm 0.3$ & $-15.95$ & N/A \\
\midrule
Doob's Seq2Seq & -- & $1506 \pm 0.5$ & $245.05 \pm 0.02$ & N/A \\
w/ Annealing & -- & $1583 \pm 0.3$ & $592.13 \pm 0.26$ & N/A \\
w/ Interpolation & -- & $1602 \pm 0.5$ & $3.46 \pm 0.03$ & N/A \\
\midrule
MaxLL & -- & $1532$ & $615$ & N/A \\
w/ Annealing & -- & $1599$ & $233$ & N/A \\
w/ Interpolation & -- & $1545$ & $619$ & N/A \\
\midrule
Chain-of-States & 2.4e5 & $-71.1$ & $-71.1$ & $-82.9$ \\
String Method & 3.3e6 & $-67.7$ & $-67.7$ & $-85.3$ \\
NEB & 7.7e6 & $-67.7$ & $-67.7$ & $-83.5$ \\
\midrule
AdaPath & 6.4e4 & $-80.26 \pm 0.20$ & $-86.58$ & $-85.1 \pm 1.6$ \\
\bottomrule
\end{tabular}
\end{table}

AdaPath requires approximately $3.8\times$ fewer energy function evaluations (6.4e4 vs. 2.4e5) compared to the best classical method (Chain-of-States) while achieving superior energy barriers ($-86.6$ vs. $-71.1$ kJ/mol for the minimum maximum energy). Compared to the Doob's Lagrangian approach, AdaPath requires approximately $800\times$ fewer energy function evaluations (6.4e4 vs. 5.1e7) while achieving significantly lower maximum energy barriers. The classical methods demonstrate substantial variation in computational efficiency, with Chain-of-States providing the best computational performance among baseline approaches in terms of function evaluations. However, AdaPath still requires fewer evaluations while achieving better energy minimization. The lower number of required evaluations for AdaPath is likely due to the efficient neural representation compared to discrete path sampling approaches used by classical methods, and the large number of images that these methods need to function well.

\section{Benchmarks: AIB9}
\label{sec:bench_aib9}

We conducted similar benchmarks on the AIB9 peptide system, a 129-atom artificial protein that exhibits well-defined metastable states and realistic conformational changes. This system serves as an intermediate complexity benchmark between simple dipeptides and larger protein systems. For the classical methods, we performed comprehensive hyperparameter optimization: Chain-of-States was evaluated across path discretizations (20 to 60 images), spring constants (1e-6 to 1e-1), and learning rates (1e-3 to 1e-1), achieving optimal performance with 40 images, spring constant 1e-5, and Adam optimizer with learning rate 1e-2. String Method was tested across image counts (20 to 60), timesteps (1 to 20), and reparameterization frequencies (1 to 15 steps), with best results using 60 images, timestep 10, and reparameterization every three steps. NEB was optimized across path discretizations (20 to 60 images), spring constants (1e-7 to 1e-1), and learning rates (1e-6 to 1e-1), yielding an optimal configuration with 60 images, spring constant 1e-3, and Adam optimizer with learning rate 1e-5.

For the Doob's Lagrangian method, we adopted the settings reported for the Chignolin system in \cite{du_doobs_2024}, as it represents a similar-sized protein system. Due to observed training instability, we added gradient clipping (1e-2) to stabilize optimization. We also found the original parameterization to be slightly under-parameterized for this system and increased the MLP width to 768 per layer across five layers. While the resulting energies appear high, they are consistent with the 3000 kJ/mol values reported for Chignolin in the original work, suggesting this energy scale is characteristic of the method's behavior on protein systems of this size.

\begin{table}[h]
\centering
\caption{Performance comparison of classical transition path methods, Doob's Lagrangian, and AdaPath on AIB9 peptide system. Results show computational efficiency, energy barriers (kJ/mol), and path quality metrics. All methods were optimized through systematic hyperparameter sweeps to ensure fair comparison. Max Energy represents the mean $\pm$ standard deviation across multiple seeds for neural methods, and the best single result for classical methods. MinMax Energy shows the lowest maximum energy barrier achieved across all runs. Mean Energy accounts for path traversal speed through different energy regions (kJ/mol).}
\label{tab:aib9_benchmark}
\begin{tabular}{lrrrr}
\toprule
\textbf{Method} & \textbf{\# Evals} & \textbf{Max Energy} & \textbf{MinMax Energy} & \textbf{Mean Energy} \\
\midrule
Chain-of-States & 2.9e6 & $-461.0$ & $-461.0$ & $-687.2$ \\
String Method & 3.1e6 & 1.3e4 & 1.3e4 & 3.2e3 \\
NEB & 4.1e6 & 2.7e3 & 2.7e3 & $-128.0$ \\
\midrule
Doob's Lagrangian & 1.2e7 & $1.6e4 \pm 7.8e3$ & 5.8e3 & N/A \\
\midrule
AdaPath & 4.7e6 & $-496 \pm 179$ & $-644$ & $-767 \pm 24$ \\
\bottomrule
\end{tabular}
\end{table}

The results demonstrate AdaPath's superior performance compared to both classical methods and the Doob's Lagrangian approach on this complex system. AdaPath achieved the best overall energy minimization, with a minimum maximum energy of $-644$ kJ/mol, substantially outperforming Chain-of-States ($-461$ kJ/mol), String Method (1.3e4 kJ/mol), NEB (2.7e3 kJ/mol), and Doob's Lagrangian (5.8e3 kJ/mol). The arc-length-adjusted mean energies further highlight AdaPath's advantages, achieving $-767 \pm 24$ kJ/mol. Notably, AdaPath demonstrated both superior energy minimization and significantly improved consistency across multiple seeds, with 3 out of 5 seeds achieving excellent results below $-590$ kJ/mol, while Doob's Lagrangian showed high variability with a standard deviation of 7.8e3 kJ/mol. The substantial energy differences between methods highlight the importance of the optimization objective and neural architecture design for effective transition path discovery.

\section{Benchmarks: BPTI}
\label{sec:bench_bpti}
We conducted benchmarks on the BPTI (Bovine Pancreatic Trypsin Inhibitor) system to evaluate our AdaPath method against classical transition path discovery approaches on a significantly larger and more complex biomolecular system, BPTI. The system setup, including explicit solvation, ionic strength, and force field parameters, is detailed in Appendix~\ref{sec:bpti_setup}.

For the classical methods (Chain-of-States, String Method, and NEB), we employed the same hyperparameter configurations that proved optimal for the AIB9 system rather than conducting comprehensive sweeps, as the computational cost of extensive hyperparameter optimization on a system of this size would be prohibitive. Specifically, we used the following methods: Chain-of-States with 40 images, a spring constant of 1e-5, and the Adam optimizer with a learning rate of 1e-2; String Method with 60 images, a timestep of 10, and reparameterization every three steps; and NEB with 60 images, a spring constant of 1e-3, and the Adam optimizer with a learning rate of 1e-5. 

For AdaPath, we used the training parameters specified in Appendix~\ref{sec:bpti_setup}, which is a more parameter-efficient variant of the model that performed best on the smaller AIB9 system.

\begin{figure}[h]
\centering
\includegraphics[width=0.8\textwidth]{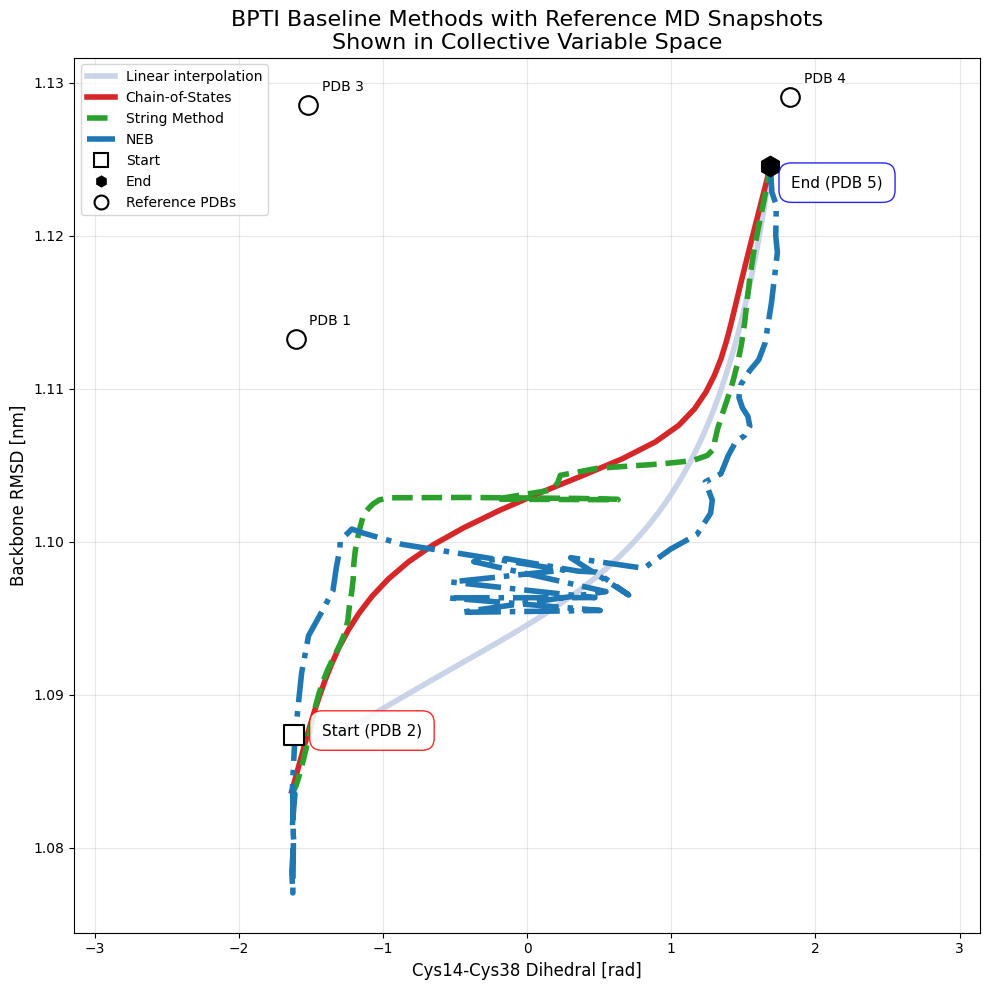}
\caption{Transition paths discovered by classical methods for the BPTI system projected onto the same collective variables used in the main paper: Cys14-Cys38 disulfide torsion angle versus backbone RMSD of residues 4-54. The Chain-of-States method (red solid line), String Method (green dashed line), and NEB (blue dash-dot line) either fail to converge and/or show erratic zigzagging behavior. Reference structures from the D.E. Shaw MD simulation are shown as open circles (PDB 1, 3, 4), with the start configuration (PDB 2) and end configuration (PDB 5) marked accordingly. Notably, none of the classical methods successfully navigate through all intermediate reference structures, in contrast to AdaPath's performance shown in the main paper.}
\label{fig:bpti_classical_paths}
\end{figure}

\begin{table}[h]
\centering
\caption{Performance comparison of classical transition path methods and AdaPath on the BPTI system with explicit solvent. Results show computational efficiency, energy barriers (kJ/mol), and path quality metrics. Max Energy shows mean $\pm$ std dev across seeds for AdaPath, best single result for classical methods.}
\label{tab:bpti_benchmark}
\begin{tabular}{lrrrr}
\toprule
\textbf{Method} & \textbf{\# Evals} & \textbf{Max Energy} & \textbf{MinMax Energy} & \textbf{Mean Energy} \\
\midrule
Chain-of-States & 1.4e6 & $-1.3e4$ & $-1.3e4$ & $-2.0e4$ \\
String Method & 3.8e6 & 1.4e4 & 1.4e4 & 4.3e3 \\
NEB & 1.1e6 & 1.6e5 & 1.6e5 & 2.3e5 \\
\midrule
AdaPath & 1.9e5 & $-4.3e4 \pm 3.1e4$ & $-6.4e4$ & $-1.2e5 \pm 2.7e4$ \\
\bottomrule
\end{tabular}
\end{table}

The results demonstrate AdaPath's superior performance on this complex, explicitly solvated system. Figure~\ref{fig:bpti_classical_paths} shows that among classical approaches, Chain-of-States produces the most reasonable pathway but fails to navigate through all intermediate reference structures from the D.E. Shaw simulation. The String Method fails to converge to a good solution, while NEB displays a jagged path, a known issue for this method in high-dimensional systems \citep{zhang_free-end_2016}. In contrast, AdaPath successfully discovers transition paths that pass through all intermediate reference structures, as demonstrated in Figure~\ref{fig:bpti_analysis}, while achieving significantly lower maximum energy barriers.
\begin{figure}[h]
\centering
\includegraphics[width=0.8\textwidth]{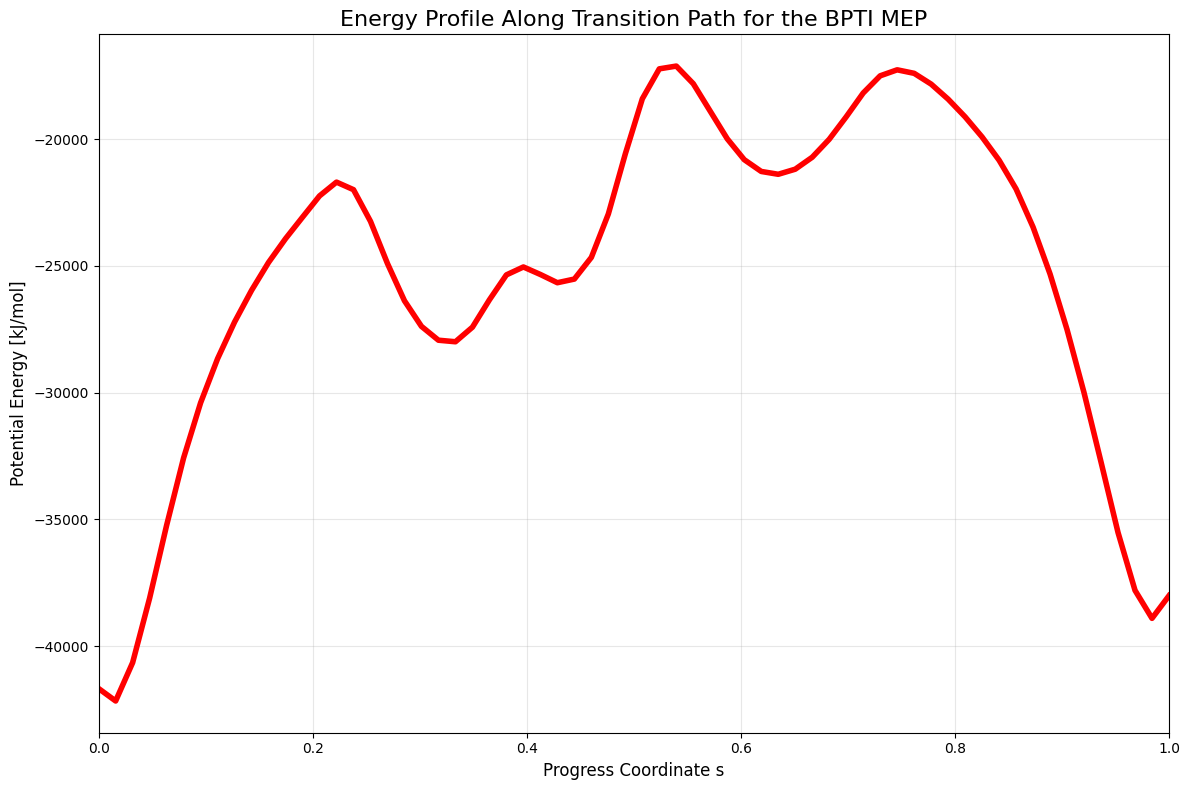}
\caption{Energy profile along the discovered BPTI transition pathway. The plot shows the potential energy landscape as a function of the progress coordinate $s \in [0,1]$ for one representative MEP connecting the two BPTI conformational states. The profile reveals multiple intermediate energy barriers and metastable states along the transition pathway.}
\label{fig:bpti_energy_profile}
\end{figure}

\section{Other related Methods}
\label{sec:comparison_with_other_methods}
While our primary benchmark comparison focuses on the Doob's Lagrangian method due to similarity in task and system setup, examining how our approach relates to other methods for transition path discovery/MEP generation is also informative. Here, we present results from two related approaches in the literature. However, we emphasize that direct numerical comparisons should be treated with caution due to differences in force fields, optimization objectives, and units used.

\begin{table}[ht]
\centering
\caption{Comparison between nudged elastic band (NEB) and implicit neural representation (INR) approaches for alanine dipeptide, adapted from \citet{ramakrishnan_implicit_2025}. The table shows the transition state (TS) energies achieved by each method after their complete optimization procedure, alongside the total number of force field evaluations required.}
\label{tab:neb_inr}
\begin{tabular}{lrr}
\toprule
\textbf{Method} & \textbf{\# Evaluations} & \textbf{Max Energy (eV)} \\
\midrule
NEB & 8.1e3 & $11.0$ \\
INR & 3.0e3 & $0.407$ \\
\bottomrule
\end{tabular}
\end{table}

\begin{table}[ht]
\centering
\caption{Computational efficiency comparison between different sampling methods for transition paths on alanine dipeptide, adapted from \citet{raja_action-minimization_2025}. The table compares traditional methods with an Onsager-Machlup (OM) optimization approach that uses a pre-trained diffusion model as a surrogate for molecular force fields. The ``CVs'' column indicates whether collective variables were required (a potential limitation).}
\label{tab:om_optimization}
\begin{tabular}{lrrr}
\toprule
\textbf{Method} & \textbf{CVs} & \textbf{\# Evaluations} & \textbf{Runtime / Path} \\
\midrule
MCMC (Two-Way Shooting) & No & $\geq$ 1e9 & $\geq$ 100 hours \\
Metadynamics & Yes & 1e6 & 10 hours \\
OM Opt. (Diffusion Model) & No & 1e4 & 50 min \\
\bottomrule
\end{tabular}
\end{table}

Table~\ref{tab:neb_inr} presents results from \citet{ramakrishnan_implicit_2025} comparing the Nudged Elastic Band (NEB) method with their Implicit Neural Representation (INR) approach for MEP discovery. Both our AdaPath approach and their INR method utilize lightweight, preexisting molecular dynamics force fields, though we employ different such force fields. Their approach requires a more complex loss function that evaluates both energy and forces along with path parameterization and its gradient (path velocity). However, it needs fewer force field evaluations than our method. It is important to note that the "Max Energy" reported in Table~\ref{tab:neb_inr} is measured in electron volts (eV) and appears to be adjusted for ground state energy reference, which makes direct energy comparisons with our method challenging. The difference in force fields used further complicates any direct energy value comparisons between methods.

In contrast, Table~\ref{tab:om_optimization} presents results from \citet{raja_action-minimization_2025}, who take a fundamentally different approach by re-purposing pre-trained generative models to perform sampling of transition paths. A key advantage of their method is that the generative model provides significantly better initial guesses for transition paths based on learned molecular distributions. Their approach requires evaluating both the pretrained model (as a surrogate force field) and its gradient for their loss function, which is more computationally intense due to the high number of parameters of the pretrained model as compared to a physics-derived force field.

These tables corroborate our findings in the Doob's Lagrangian comparison: methods that optimize for a single MEP consistently require orders of magnitude fewer force field evaluations than approaches that characterize the complete transition path ensemble.

\subsection{Connection between Doob's Lagrangian and MEP Optimization}
\label{sec:doob}

We elaborate on the method of Doob's h-transform for sampling transition paths \citep{du_doobs_2024}. This explanation is relevant as both methods employ neural networks to represent molecular state transitions, either as a minimum-energy path in our case or the complete transition path ensemble in theirs. Notably, the parametrization of the mean trajectory in Du et al. resembles our MEP representation, differing primarily in our use of blending functions and spatial embeddings for the neural network component.

Doob's Lagrangian formulation considers the following action functional:
\begin{equation}
S = \min_{q,v} \int_0^T dt \int dx \, q_{t|0,T}(x)\langle v_{t|0,T}(x), G_t v_{t|0,T}(x)\rangle
\end{equation}

This functional quantifies the cost of controlling a stochastic process to achieve desired endpoint conditions, subject to the system's energy landscape. Intuitively, it measures the amount of "effort"/action needed to steer the dynamics from starting state A to target state B, with smaller values indicating more probable transition paths.

In this formulation, $q_{t|0,T}(x)$ is the probability density at time $t$ given boundary conditions, parameterized as a Gaussian $\mathcal{N}(x | \mu_{t|0,T}, \Sigma_{t|0,T})$, and $v_{t|0,T}(x)$ is the control vector field satisfying $v_{t|0,T}(x) = \frac{1}{2}(G_t)^{-1}(u_{t|0,T}(x) - b_t(x))$, with $b_t(x) = -\nabla V(x)$ representing the reference drift for overdamped dynamics.

This action functional describes a Wasserstein Lagrangian flow as formalized in \citep{neklyudov_action_2023} and \citep{neklyudov_computational_2024}. In this framework, one can learn stochastic dynamics from samples, thereby learning a process that defines a time-dependent density evolving from an initial to a final state. 

The difference with our approach is that we minimize the action of a single deterministic path rather than a complete distribution of paths. This simplification is reflected in our loss function, which contains one fewer expectation than the full Doob's Lagrangian.

\section{Ablation Studies}
\label{sec:ablation}

This section presents a series of ablation studies examining the impact of various components of our AdaPath approach. These experiments help validate our design choices and provide insights into the importance of different architectural and training elements. We conduct these ablation studies on the AIB9 system because Alanine dipeptide, as a benchmark system, is too simple to provide meaningful insights into the performance of our method. For each ablation study, we report the lowest maximum energy of the path among all training runs, the mean and standard deviation of the arc-length adjusted energy, and the mean number of iterations to reach the lowest maximum energy. We run each setting in the ablations 5 times. We use a batch size of 1, where each batch contains 16 values of the progress coordinate $s$ for most cases.

\subsection{General Settings for Ablation Studies}
\label{sec:ablation_general_settings}

All ablation studies are conducted using the following base configuration, with each study varying only specific parameters while keeping all others fixed:

\textbf{Molecular System Parameters:}
\begin{itemize}
    \item Molecular system: AIB9 peptide with 129 atoms
    \item Force field: AMBER ff15ipq-m for protein mimetics
    \item Cutoff method: No cutoff applied
    \item Simulation box size: 10.0 nm
    \item Neighbor list cutoff distance: 4.0 nm
\end{itemize}

\textbf{Training Configuration:}
\begin{itemize}
    \item Number of iterations: 100,000 per run
    \item Number of independent runs: 5 per parameter setting
    \item Evaluation frequency: Every 2000 training steps
    \item Optimizer: Muon as implemented in Optax.
    \item Learning rate: 1e-2
\end{itemize}

\textbf{Path Sampling Parameters:}
\begin{itemize}
    \item Batch size: 1
    \item Frames per batch: 16 points sampled along the progress coordinate $s$
    \item Interpolation method: Linear
\end{itemize}

\textbf{Network Architecture (Base Configuration):}
\begin{itemize}
    \item Network depth: 6 MLP blocks.
    \item MLP Expansion Factor: 4, meaning that each MLP block, consisting of two linear layers, has four times the hidden dimensionality as input and output dimensionality.
    \item Atom Embedding Dimension: 32.
    \item Time embedding: Sinusoidal with maximum period of $1$
    \item Activation function: GeLu throughout the network
    \item Conditioning MLP settings: 3 Layers, with a hidden dimension of 1024 and GeLu activations.
\end{itemize}

\subsection{Architecture Comparison: AdaPath vs Standard MLP}
To validate our core claim that parameter sharing across atoms provides superior scalability, we compare AdaPath against a standard MLP baseline that directly maps progress coordinate $s$ to the complete molecular configuration without parameter sharing. This comparison is conducted across all three molecular systems (Alanine dipeptide, AIB9, and BPTI) to demonstrate scaling behavior.

The standard MLP baseline uses a conventional feedforward architecture with six hidden layers of width determined by 4 times the number of atoms in the system. This was found to be suitable for the two small systems. Intuitively, this aligns well with deep learning's collective wisdom of having a higher hidden dimension than the input and output dimensions of the network. We only lower this to one time the atom count for the BPTI system, as this already leads to an MLP with 1= billion parameters, as more was not runnable for our hardware. The network takes sinusoidal embeddings of the progress coordinate $s$ as input and directly outputs the $3N$-dimensional molecular configuration, where $N$ is the number of atoms in the system. The standard MLP (TransitionMLP) uses the same interpolation strategy and endpoint constraints as AdaPath but processes all atomic coordinates through shared dense layers rather than using per-atom embeddings with parameter sharing.

All other training parameters remain identical to the base configuration described in Section~\ref{sec:ablation_general_settings}, with both architectures using the same learning rate of 1e-2 with the Muon optimizer for the two smaller systems and the Nadamw optimizer for BPTI with the learning rate scheduling as outlined in \ref{sec:bpti_setup}.

\begin{table}[h]
\centering
\caption{Architecture comparison between AdaPath and standard MLP on Alanine dipeptide system. Results demonstrate the parameter efficiency and performance of the parameter-sharing approach on this simple benchmark system.}
\label{tab:ablation_arch_alanine}
\begin{tabular}{lrrrr}
\toprule
\textbf{Architecture} & \textbf{Parameters} & \textbf{Max Energy} & \textbf{MinMax Energy} & \textbf{Mean Energy} \\
\midrule
Standard MLP & 5e4 & $5 \pm 160$ & $-68.1$ & $-25 \pm 103$ \\
AdaPath & 2e6 & $-80.3 \pm 0.2$ & $-86.6$ & $-85.1 \pm 1.6$ \\
\bottomrule
\end{tabular}
\end{table}

\begin{table}[h]
\centering
\caption{Architecture comparison between AdaPath and standard MLP on AIB9 peptide system. The 129-atom system provides a more challenging test case for evaluating architectural scaling behavior.}
\label{tab:ablation_arch_aib9}
\begin{tabular}{lrrrr}
\toprule
\textbf{Architecture} & \textbf{Parameters} & \textbf{Max Energy} & \textbf{MinMax Energy} & \textbf{Mean Energy} \\
\midrule
Standard MLP & 2e6 & $4.3e4 \pm 9.3e3$ & $2.99e4$ & $1.92e4 \pm 2.3e3$ \\
AdaPath & 2e6 & $-496 \pm 179$ & $-644$ & $-767 \pm 24$ \\
\bottomrule
\end{tabular}
\end{table}

\begin{table}[h]
\centering
\caption{Architecture comparison between AdaPath and standard MLP on the BPTI system with explicit solvent. This large system provides the most stringent test of architectural scalability claims.}
\label{tab:ablation_arch_bpti}
\begin{tabular}{lrrrr}
\toprule
\textbf{Architecture} & \textbf{Parameters} & \textbf{Max Energy} & \textbf{MinMax Energy} & \textbf{Mean Energy} \\
\midrule
Standard MLP & 1.4e9 & $2.2e4 \pm 2.0e4$ & $2.98e3$ & $1.92e4 \pm 1.2e4$ \\
AdaPath & 3.5e6 & $-4.3e4 \pm 3.1e4$ & $-6.4e4$ & $-1.15e5 \pm 2.7e4$ \\
\bottomrule
\end{tabular}
\end{table}

The results demonstrate AdaPath's superior performance and scalability across system sizes. The AdaPath's parameter count is dominated by the progress coordinate MLP that outputs the adaLn parameters. This produces a baseline parameter count that puts the AdaPath architecture above the MLP for small systems, but for larger ones, this gap inverts sharply. 

On the alanine dipeptide system, AdaPath achieves significantly lower and more consistent maximum energies (-86.6 vs -68.1 kJ/mol for minimum maximum energy) with substantially reduced variance. The performance gap becomes dramatic on the larger AIB9 system, where the standard MLP achieves a best maximum energy of 2.99e4 kJ/mol compared to AdaPath's -644 kJ/mol, representing a 30,000+ kJ/mol improvement in path quality. 
\subsection{Optimizer Comparison: Path Continuity Analysis}

To validate our claim that the Muon optimizer maintains better path continuity compared to Adam for small systems, we conduct a comparative analysis using path velocity metrics over training iterations. We train identical AdaPath networks on the AIB9 system using both optimizers with learning rates of 1e-2 for Muon and 1e-4 for Adam, respectively. All other training parameters remain identical to the base configuration.

We track the maximum path velocity over training iterations, computed as $\max_{s \in [0,1]} \|\frac{d\varphi_\theta(s)}{ds}\|$, where high velocities indicate rapid changes in molecular configuration that can lead to discontinuous "teleportation" between states rather than smooth transitions.

\begin{figure}[h]
\centering
\includegraphics[width=0.8\textwidth]{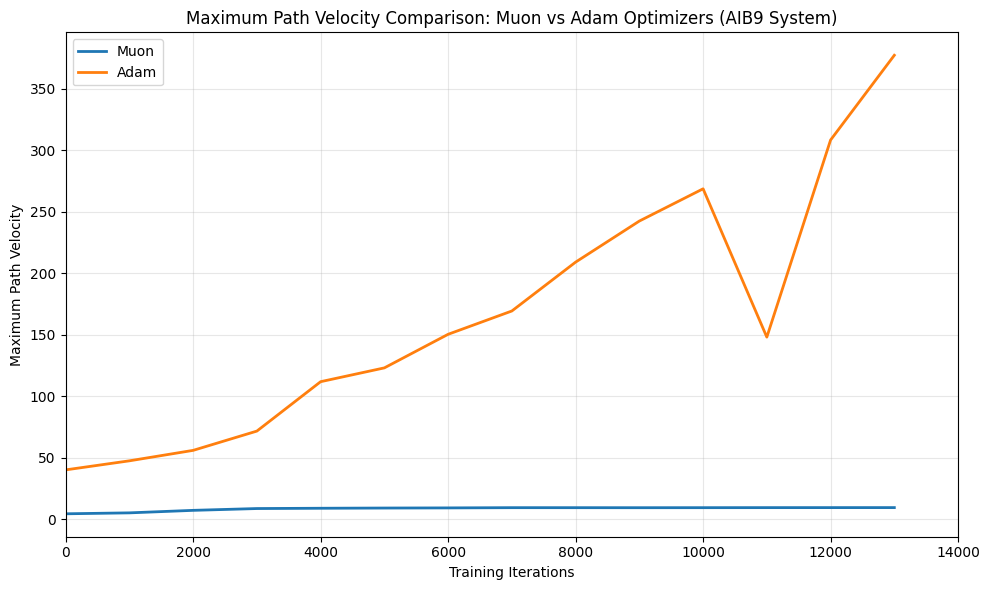}
\caption{Maximum path velocity over training iterations comparing Muon and Adam optimizers on the AIB9 system. Path velocity is computed as the maximum spatial derivative along the transition path at each training step. High velocities indicate rapid configurational changes that can lead to discontinuous transitions, while moderate velocities suggest smoother, more physically meaningful pathways.}
\label{fig:path_velocity_comparison}
\end{figure}

\begin{table}[h]
\centering
\caption{Optimizer comparison between optimizers on AIB9 system.}
\label{tab:optimizer_comparison}
\begin{tabular}{lrr}
\toprule
\textbf{Optimizer} & \textbf{Max Energy} & \textbf{Mean Energy} \\
\midrule
Adam & $3.0e3$ & $-802$ \\
Muon & $-496 \pm 179$ & $-767 \pm 24$ \\
\bottomrule
\end{tabular}
\end{table}

The Adam-trained system exhibits extremely high path velocities, indicating that it rapidly switches from one state to another without passing through the transition region. This is corroborated by the low mean but high max energies.

\subsection{Energy Transformation Analysis}

We evaluate the impact of our logarithmic energy transformation on training stability by comparing performance with and without the log transformation applied to the potential energy in the loss function. The logarithmic transformation is applied as $\mathcal{L}_{\text{log}}(\theta) = \frac{1}{B}\sum_{j=1}^{B} \log(U(\varphi_\theta(s_j)) + c)$ where $c = 0.1$ ensures positivity, while the baseline uses the direct energy $\mathcal{L}_{\text{direct}}(\theta) = \frac{1}{B}\sum_{j=1}^{B} U(\varphi_\theta(s_j))$. All other training parameters remain identical to the base configuration.

\begin{table}[h]
\centering
\caption{The impact of logarithmic energy transformation on AIB9 system performance and training stability.}
\label{tab:energy_transformation}
\begin{tabular}{lrrr}
\toprule
\textbf{Energy Transform} & \textbf{Max Energy} & \textbf{MinMax Energy} & \textbf{Mean Energy} \\
\midrule
Direct Energy & $3.3e4 \pm 1.7e4$ & $3.0e3$ & $6.4e3 \pm 3.1e3$ \\
Log Transform & $-496 \pm 179$ & $-644$ & $-767 \pm 24$ \\
\bottomrule
\end{tabular}
\end{table}

The logarithmic transformation achieves significantly lower energies with reduced variance, demonstrating improved training stability and path quality over direct energy optimization.
\subsubsection{Reference Trajectory Generation}

To generate reference conformational data, we performed an extended molecular dynamics simulation using OpenMM with the following parameters:

\begin{itemize}
    \item \textbf{Force field}: AMBER ff15ipq-m
    \item \textbf{Temperature}: 500 K (elevated to enhance sampling of conformational transitions)
    \item \textbf{Integrator}: Langevin Middle Integrator with friction coefficient of 1 ps$^{-1}$
    \item \textbf{Time step}: 1 femtosecond
    \item \textbf{Non-bonded treatment}: No cutoff (NoCutoff method)
    \item \textbf{Constraints}: Hydrogen bonds constrained using the HBonds method
    \item \textbf{Total simulation time}: 900 nanoseconds
    \item \textbf{Output frequency}: Coordinates saved every 1000 steps (1 ps intervals)
\end{itemize}

The simulation protocol consisted of initial energy minimization followed by velocity assignment at the target temperature, a brief 10 ps equilibration period, and then the production run. The elevated temperature of 500 K was chosen to accelerate conformational sampling and ensure adequate transitions between metastable states during the simulation timeframe.

\subsubsection{State Definition and Analysis}

To characterize the conformational landscape, we focused on the backbone dihedral angles of the central residues. Specifically, we computed the $\phi$ and $\psi$ dihedral angles for residues at indices [56, 58, 60, 69] and [58, 60, 69, 71], respectively, corresponding to the central amino acids of the peptide chain.

Two metastable states were defined using ellipsoidal boundaries in the $\phi$-$\psi$ dihedral space:
\begin{align}
\text{State A}: &\quad \left(\frac{\phi + 1.0}{0.1}\right)^2 + \left(\frac{\psi + 0.45}{0.1}\right)^2 \leq 1 \\
\text{State B}: &\quad \left(\frac{\phi - 1.0}{0.1}\right)^2 + \left(\frac{\psi - 0.45}{0.1}\right)^2 \leq 1
\end{align}

These elliptical regions capture the distinct conformational basins observed in the free energy landscape projected onto the central residue dihedral angles.

\subsubsection{Neural Network Training Setup}

For the neural network-based MEP optimization, we employed a different force field setup optimized for differentiable molecular dynamics:

\begin{itemize}
    \item \textbf{Force field}: AMBER ff15ipq-m for protein mimetics, implemented through DMFF
    \item \textbf{Non-bonded treatment}: No cutoff method for consistency with the reference simulation
    \item \textbf{Simulation box}: 11.0 nm cubic box
    \item \textbf{Neighbor list cutoff}: 5.0 nm
\end{itemize}

\subsubsection{Start- and End-state Setup}

Start and end configurations for MEP optimization were randomly sampled from the reference trajectory frames that satisfied the respective state definitions. The Kabsch algorithm was employed to optimally align the end configuration to the start configuration, removing translational and rotational degrees of freedom. This alignment procedure ensures that the neural network focuses on learning the internal conformational changes rather than rigid-body motions, leading to more efficient and physically meaningful transition path discovery.

\section{BPTI System Preparation}
\label{sec:bpti_setup}
As reference conformations, we used the structures from the D.E. Shaw Research BPTI simulation of BPTI \citep{shaw_atomic-level_2010}. The initial and final states for our MEP optimization were selected as the second snapshot (index 1) and the second-to-last snapshot (index -2) from the provided reference structures, as they were the most separated in the CVs determined in the original study, named the backbone RMSD and the disulfide torsion angle.

\subsection{Explicit Solvent System Preparation}

For the BPTI system, we developed an approach to handle explicit water molecules and ions using periodic boundary conditions:

\begin{itemize}
    \item \textbf{System solvation}: Both start and end protein configurations were placed in 3.5 nm cubic periodic boxes and solvated with TIP3P water molecules using OpenMM's Modeller class. The system included Na$^+$ and Cl$^-$ ions at 0.04 M ionic strength to approximate physiological conditions.
    
    \item \textbf{Energy minimization}: To preserve the reference protein conformations from the D.E. Shaw simulation while allowing solvent relaxation, all protein heavy atoms were fixed by setting their masses to zero during energy minimization. This completely freezes the protein structure while allowing water molecules and ions to relax.
    
    \item \textbf{Molecule matching}: Since the endpoint configurations were solvated independently, we equalized atom counts by randomly removing excess molecules, then used the Hungarian algorithm to optimally pair water molecules and ions between configurations based on center-of-mass positions. This matching minimizes solvent displacement during transitions.
\end{itemize}

The resulting system contained approximately 3,500 atoms total in a periodic cubic box, maintaining proper solvation while using standard periodic boundary conditions compatible with conventional MD force fields.

\subsection{Training Parameters}

The BPTI MEP optimization employed a two-stage training strategy due to spiking loss late during training. This is likely due to sharp minima in large molecular configurations, including water, where small changes in atomic coordinates can produce drastic energy spikes. This necessitated transitioning to more conservative learning parameters in Stage 2 to maintain training stability.

\textbf{Initial Training, iterations 0-6,000:}
\begin{itemize}
    \item Learning rate: $1 \times 10^{-4}$
    \item Gradient clipping threshold: $5 \times 10^{-4}$
\end{itemize}

\textbf{Fine-tuning, iterations 6,000-12,000:}
\begin{itemize}
    \item Learning rate: $1 \times 10^{-6}$
    \item Gradient clipping threshold: $1 \times 10^{-4}$
\end{itemize}

\textbf{Architecture and General Training Parameters:}
\begin{itemize}
    \item MLP Expansion Factor: 2
    \item Layers: 5
    \item Hidden dimension: 64
    \item Progress embedding dimension: 1024
    \item Conditioning depth: 3 layers
    \item Conditioning hidden size: 1024
    \item Optimizer: NAdam with weight decay (nadamw)
    \item Weight decay: $1 \times 10^{-3}$
    \item Beta parameters: $\beta_1 = 0.9$, $\beta_2 = 0.99$, $\epsilon = 1 \times 10^{-8}$
    \item Batch size: 1
    \item Frames per batch: 16
    \item Total iterations: 12,000
    \item Evaluation frequency: Every 1000 iterations
\end{itemize}

\subsection{Force Field Parameters}

We used the following force field parameters to match the D.E. Shaw simulation conditions:
\begin{itemize}
    \item \textbf{Protein force field}: AMBER99sb-ILDN (amber99sbildn.xml)
    \item \textbf{Water force field}: TIP3P (tip3p.xml)
\end{itemize}

\section{Training Stability}
\label{sec:energyspikes}

Atomistic force fields present unique optimization challenges due to the presence of Lennard-Jones interactions, which are typically modeled using the 12-6 potential:
\begin{align}
V_{\text{LJ}}(r_{ij}) = 4\epsilon_{ij}\left[\left(\frac{\sigma_{ij}}{r_{ij}}\right)^{12} - \left(\frac{\sigma_{ij}}{r_{ij}}\right)^{6}\right]
\end{align}
where $r_{ij}$ is the distance between atoms $i$ and $j$, $\epsilon_{ij}$ is the well depth, and $\sigma_{ij}$ is the collision diameter. The $r^{-12}$ repulsive term creates extremely steep energy barriers when atoms approach each other too closely, leading to numerical instabilities during neural network training. As $r_{ij} \rightarrow 0$, the potential energy diverges to infinity, causing gradient magnitudes that can destabilize the optimization process.

To mitigate these issues, we employ a two-pronged approach. First, we apply a logarithmic transformation to the potential energy after shifting it to ensure positivity:
\begin{align}
\mathcal{L}_{\text{transformed}}(\theta) = \frac{1}{B}\sum_{j=1}^{B} \log(U(\phi_\theta(s_j)) + c)
\end{align}
Where $c$ is a positive constant to ensure the argument of the logarithm remains positive. This transformation preserves the location of energy minima while significantly compressing the magnitude of high-energy spikes. The monotonic nature of the logarithm ensures that $\arg\min_\theta \mathcal{L}_{\text{transformed}}(\theta) = \arg\min_\theta \mathcal{L}(\theta)$, preserving the optimization objective. However, the model now places different weights on different parts of the path than before. In practice, we find that this drastically improves training stability.

Second, we implement gradient norm-based gradient clipping to handle any remaining numerical instabilities. This combination of logarithmic energy transformation and gradient clipping provides robust training stability while maintaining the physical validity of the discovered minimum-energy paths.

\subsection{Logarithmic Energy Transformation Perturbation Analysis}
\label{sec:log-transform-justification}
Consider the true minimum energy path $\varphi^*(s)$ for $s \in [0,1]$ satisfying:
\begin{itemize}
    \item Boundary conditions: $\varphi^*(0) = x_A$ and $\varphi^*(1) = x_B$
    \item Orthogonality condition: $[\nabla U(\varphi^*(s))]^{\perp} = \nabla U - (\nabla U \cdot \hat{\tau})\hat{\tau} = 0$
\end{itemize}
where $\hat{\tau} = \frac{d\varphi^*/ds}{|d\varphi^*/ds|}$ is the unit tangent vector along the path.

Consider a perturbed path $\varphi(s) = \varphi^*(s) + \epsilon \delta(s)$ where $\epsilon \ll 1$ and $\delta(0) = \delta(1) = 0$ (fixed endpoints). At each point, decompose the perturbation:
\begin{align}
    \delta(s) = \delta_{\parallel}(s) + \delta_{\perp}(s)
\end{align}
where $\delta_{\parallel} = (\delta \cdot \hat{\tau})\hat{\tau}$ and $\delta_{\perp} \cdot \hat{\tau} = 0$.

Under perturbation, the energy at each point becomes:
\begin{align}
    U(\varphi^* + \epsilon\delta) = U(\varphi^*) + \epsilon \nabla U(\varphi^*) \cdot \delta + \frac{\epsilon^2}{2} \delta^T H \delta + O(\epsilon^3)
\end{align}
where $H$ is the Hessian of $U$, and we examine the perpendicular components of the perturbation and how they interact with the Hessian, as these are the components of interest in this context. Since $[\nabla U(\varphi^*)]^{\perp} = 0$ at the MEP:
\begin{align}
    U(\varphi^* + \epsilon\delta) = U(\varphi^*) + \epsilon[\nabla U]_{\parallel} \cdot \delta_{\parallel} + \frac{\epsilon^2}{2} \delta_{\perp}^T H_{\perp} \delta_{\perp} + O(\epsilon^3)
\end{align}

The change in the discretized original loss:
\begin{align}
    \Delta\mathcal{L} = \frac{1}{B}\sum_{j=1}^B [U(\varphi^*(s_j) + \epsilon\delta(s_j)) - U(\varphi^*(s_j))]
\end{align}
Substituting the energy expansion:
\begin{align}
    \Delta\mathcal{L} = \frac{\epsilon}{B}\sum_{j=1}^B [\nabla U]_{\parallel} \cdot \delta_{\parallel}(s_j) + \frac{\epsilon^2}{2B}\sum_{j=1}^B \delta_{\perp}^T(s_j) H_{\perp} \delta_{\perp}(s_j)
\end{align}

For the log-transformed loss:
\begin{align}
    \Delta\mathcal{L}_{\log} &= \frac{1}{B}\sum_{j=1}^B [\log(U(\varphi^* + \epsilon\delta) + c) - \log(U(\varphi^*) + c)]\\
    &= \frac{1}{B}\sum_{j=1}^B \log\left(1 + \frac{\epsilon[\nabla U]_{\parallel} \cdot \delta_{\parallel} + \frac{\epsilon^2}{2}\delta_{\perp}^T H_{\perp} \delta_{\perp}}{U(\varphi^*) + c}\right)
\end{align}
Using $\log(1 + x) \approx x$ for small $x$:
\begin{align}
    \Delta\mathcal{L}_{\log} = \frac{\epsilon}{B}\sum_{j=1}^B \frac{[\nabla U]_{\parallel} \cdot \delta_{\parallel}(s_j)}{U(\varphi^*(s_j)) + c} + \frac{\epsilon^2}{2B}\sum_{j=1}^B \frac{\delta_{\perp}^T(s_j) H_{\perp} \delta_{\perp}(s_j)}{U(\varphi^*(s_j)) + c}
\end{align}

Both losses exhibit:
\begin{enumerate}
    \item No first-order response to perpendicular perturbations: The $\delta_{\perp}$ terms appear only at second order
    \item Positive-definite second-order response: Since the MEP follows energy valleys, $H_{\perp} > 0$
    \item Restoring forces: Both $\Delta\mathcal{L} > 0$ and $\Delta\mathcal{L}_{\log} > 0$ for any $\delta_{\perp} \neq 0$
\end{enumerate}

The difference lies only in the weighting factor $\frac{1}{U(\varphi^*) + c}$, which scales the magnitude but not the direction of the restoring force. Therefore, both optimizations converge to the same geometric path.

\subsubsection{Parameterization and Path Stretching}

While both losses converge to the same MEP geometry, they differ in how they parameterize the path. The parallel component of the perturbation affects the path's parameterization speed $v(s) = |d\varphi/ds|$.

The gradient with respect to shifting a sample point along the path:
\begin{align}
    \frac{\partial \mathcal{L}}{\partial s_j} = \nabla U(\varphi(s_j)) \cdot \frac{d\varphi}{ds}\bigg|_{s_j} = |\nabla U|_{\parallel} \cdot v(s_j)
\end{align}
This creates forces that push sample points away from high-energy regions, potentially leading to extreme stretching where $v(s) \to \infty$ at transition states.

The corresponding gradient for the log loss:
\begin{align}
    \frac{\partial \mathcal{L}_{\log}}{\partial s_j} = \frac{1}{U(\varphi(s_j)) + c} |\nabla U|_{\parallel} \cdot v(s_j)
\end{align}

The weighting factor $\frac{1}{U + c}$ provides implicit regularization:
\begin{itemize}
    \item At transition states (high $U$): The factor is small, reducing the incentive to stretch
    \item At stable states (low $U$): The factor is larger but bounded
\end{itemize}
\section{Computational Performance}
\label{sec:performance}

For the BPTI system with explicit solvent (approximately 3,500 atoms total):
\begin{itemize}
    \item \textbf{Training time}: Approximately 15 minutes on a single NVIDIA A6000 GPU
    \item \textbf{Force field evaluations}: Approximately 192,000 evaluations (30,000 iterations x 16 frames per batch)
    \item \textbf{Trajectory generation}: 64 frames for the final MEP visualization and analysis
\end{itemize}

This computational effort is approximately six orders of magnitude less, in node-hours, than the original D.E. Shaw Research simulation on their specialized Anton supercomputer, which required approximately 1.3 million node-hours to generate the 1.03 millisecond trajectory capturing the same conformational changes.

\section{Computational Resources}
All experiments reported in this paper were conducted on a single NVIDIA A6000 GPU with 48GB of memory. Additional computational resources were used during the development phase of this project for hyperparameter tuning across network architectures and optimization parameters.

\end{document}